# Is the molecular KS relationship universal down to low metallicities?

David J. Whitworth,[1]* Rowan J. Smith,[1] Robin Tress,[2] Scott T. Kay,[1] Simon C. O. Glover,[2] Mattia C. Sormani,[2] Ralf S. Klessen[2,3]

[1]*Jodrell Bank Centre for Astrophysics, Department of Physics and Astronomy, University of Manchester, Oxford Road, Manchester M13 9PL, UK*
[2]*Universitat Heidelberg, Zentrum fur Astronomie, Institut fur Theoretische Astrophysik, Albert-Ueberle-Str. 2, 69120 Heidelberg, Germany*
[3] *Universität Heidelberg, Interdisziplinäres Zentrum für Wissenschaftliches Rechnen, Im Neuenheimer Feld 205, D-69120 Heidelberg, Germany*

**ABSTRACT**

In recent years it has been speculated that in extreme low metallicity galactic environments, stars form in regions that lack $H_2$. In this paper we investigate how changing the metallicity and UV-field strength of a galaxy affects the star formation within, and the molecular gas Kennicutt-Schmidt relation. Using extremely high resolution AREPO simulations of isolated dwarf galaxies, we independently vary the metallicity and UV-field to between 1% and 10% solar neighbourhood values. We include a non-equilibrium, time-dependant chemical network to model the molecular composition of the ISM, and include the effects of gas shielding from an ambient UV field. Crucially our simulations directly model the gravitational collapse of gas into star-forming clumps and cores and their subsequent accretion using sink particles. In this first publication we find that reducing the metallicity and UV-field by a factor of 10 has no effect on star formation, and minimal effect on the cold, dense star forming gas. The cold gas depletion times are almost an order of magnitude longer than the molecular gas depletion time due to the presence of star formation in HI dominated cold gas. We study the $H_2$ Kennicutt-Schmidt relationship that arises naturally within the simulations and find a near linear power law index of $N = 1.09 \pm 0.014$ in our fiducial 10% solar metallicity model. As the metallicity and UV-field are reduced this becomes moderately steeper, with a slope of $N = 1.24 \pm 0.022$ for our 1% solar metallicity and 1% solar UV field model.

**Key words:** galaxies: ISM – ISM: clouds – ISM: structure – hydrodynamics – stars: formation

## 1 INTRODUCTION

Dwarf galaxies are the most numerous type of galaxies in the universe and are important building blocks to assemble larger galaxies. They have low surface brightness, shallow gravitational potential wells, and relatively low star formation rates (SFRs) compared to larger galaxies. Due to their shallow gravitational potential wells, they react sensitively to external and internal processes such as feedback by supernovae (SNe) and cosmic-ray ionisation, and are therefore ideal objects to study galaxy evolution. With dwarf galaxies showing lower metallicities ($Z$), they are useful test laboratories for studying how the ISM responds to varying chemical and physical processes which are hard to model accurately for Milky Way like galaxies.

Dwarfs can be thought of as analogous to high redshift galaxies. Understanding how star formation behaves in these extreme conditions will help us to understand how the first galaxies and stars formed in the early, post reionisation, universe. Dwarf galaxies are also progenitors to larger galaxies through mergers, so an understanding of dwarfs can lead to a greater understanding of larger galaxies and their evolution. For example Shi et al. (2014) show that in nearby extremely metal poor galaxies (XMPs) star formation is inefficient. These XMPs are our closest analogous galaxies to high redshift galaxies and provide the best observational results to draw from.

At extremely low metallicities $H_2$ formation is inefficient due to the reduced availability of the dust grain surfaces that allow the HI to $H_2$ reaction to occur (Palla et al. 1983; Omukai et al. 2010). This has implications for our understanding of how stars form within these galaxies. Observations of star-forming spiral galaxies from multiple surveys (e.g. THINGS (Walter et al. 2008), SINGS (Kennicutt et al. 2003), GALEX NGS (Gil de Paz et al. 2007), the HERACLES (Leroy et al. 2008) or BIMA-SONG (Helfer et al. 2003) CO surveys and the spatially resolved PHANGS survey (Pessa et al. 2021)) have shown that the star formation surface density ($\Sigma_{\rm SFR}$) is strongly correlated with molecular hydrogen surface density ($\Sigma_{H_2}$), and only loosely correlated to atomic hydrogen or total gas surface density. The seminal work by Bigiel et al. (2008), for example, found a molecular gas Kennicutt-Schmidt relationship ($\Sigma_{\rm SFR} \propto \Sigma_{\rm H_2}^N$) with an exponent of $N = 1.0 \pm 0.2$ in spiral galaxies where the total gas relation varies dramatically from a low of $N = 1.1$ to $N = 2.7$.

---
* E-mail: david.whitworth@manchester.ac.uk

Such work is now being extended to the low metallicity regime. de los Reyes & Kennicutt (2019) looked at the relation between $H_2$ and star formation in a large sample of 307 quiescent spiral, dwarf or low-surface-brightness galaxies to revise the KS relationship. They find a revised Kennicutt-Schmidt relationship using HI + $H_2$ surface density of N= $1.4 \pm 0.07$ for spirals, which drops to N= $1.26 \pm 0.08$ for dwarfs. When looking only at $H_2$ in spirals this becomes a sublinear relationship with an exponent of N= $0.67 \pm 0.05$, significantly shallower than in the higher metallicity spirals of Bigiel et al. (2008). One of the problems with finding the molecular KS relationship for dwarf galaxies is the difficulty in detecting $H_2$ within them, Roychowdhury et al. (2015) and Filho et al. (2016) both look at the dominant HI component in CO-dark dwarf galaxies and note that it is not a good tracer for $H_2$ or star formation and the respective KS relationship.

However, characterising such relationships is always challenging in low metallicity systems due to the necessity of using an assumed $X_{\rm CO}$ conversion factor (Bolatto et al. 2013) to convert from CO emission to the actual molecular gas mass. Detecting CO emission from faint sources and constraining the $X_{\rm CO}$ factor is much harder in such systems. For these reasons numerical simulations have an important role to play in characterising star formation and its relation to molecular gas in dwarf systems.

The link between molecular gas and star formation has been a topic of some discussion in the literature. Krumholz (2012) proposed that star formation will occur in the cold atomic phase of the ISM in galaxies and clouds with metallicities of a few percent of $Z_\odot$ or lower. Similarly, Glover & Clark (2012a,b) and Hu et al. (2021) propose that molecular gas and star formation coincide in metal-rich galaxies due to gravitational instability creating favourable conditions for the formation of both, and not because of a causal connection between $H_2$ and star formation. They suggest that at low metallicities, the link between $H_2$ and star formation may break down. However these works were not full galaxy simulations, and to fully understand these ideas an analysis on galactic scales is needed.

A related issue is the effect of varying the UV field strength ($G_0$) upon star formation. Hu et al. (2016, 2017) performed simulations at a fixed metallicity of Z = 0.1 solar and found that modifying the UV field had little impact upon the star formation rate surface density. On the other hand, Ostriker et al. (2010) find a linear connection between SFR surface density and $G_0$ in gravitationally-bound clouds.

In this work we aim to investigate how changing the metallicity and average UV field impacts upon the star formation rate and resulting molecular gas KS relations in a set of high resolution isolated dwarf galaxy simulations. We select our metallicity regime such that we include dwarfs with Z = 0.01 solar where Krumholz (2012) suggested star formation becomes associated with cold atomic gas. The models include hydrodynamics, self-gravity, non-equilibrium heating, cooling and chemistry, a new chemical network suited to diffuse gas, shielding, and supernovae feedback. Crucially our simulations directly model the gravitational collapse of gas into star-forming clumps and cores and their subsequent accretion using sink particles (Bate et al. 1995; Tress et al. 2020). This is in contrast to more commonly-adopted stochastic star formation models that randomly create stars from gas above a given density given a star formation rate per freefall time. This enables us to directly link the mass involved in star formation to the energetic properties of the region of the ISM in which it forms.

The paper is structured as follows. In section 2 we lay out the numerical models used. In section 3 we detail the results from the simulations. Section 4 discusses the implications of the results and caveats. In section 5 we provide a summary of our work.

## 2 METHODS

### 2.1 Modelling

The simulations in this work are performed using the quasi-Lagrangian moving mesh code AREPO (Springel 2010). AREPO solves the hydrodynamic equations on a 3D moving-mesh with a second-order Godunov scheme that uses an exact Riemann solver using Heuns method, as described in (Pakmor et al. 2016).

To do this it uses a set of discrete points to generate a Delaunay triangulation which is then used to construct a 3D unstructured Voronoi mesh. The code allows the points to move with the local velocity of the gas, reconstructing the mesh after each time-step. This allows the cells to move, adapt and continuously change whilst the mass in each cell remains approximately constant. It also means that the fluid density has a strong influence on the resolution of the code. At higher densities the cells of the mesh are smaller, giving finer resolution; at lower densities, like the edge of the simulation, where fine resolution is not needed the cells are much larger.

AREPO uses a Barnes-Hut tree-based approach based on that of GADGET-3 (Springel 2005) to calculate the gravitational potential and can consider multiple cell/particles types. In our simulations we include both dark matter particles to model the galaxy halo, gas cells that interact hydrodynamically, collisionless star particles, and sink particles representing regions of ongoing star formation. In the following sections we discuss in more detail our custom ISM physics modules, but for a more detailed discussion of the AREPO code in general see Springel (2010) and Weinberger et al. (2019).

### 2.2 Chemical network

In our simulations we use a slightly modified version of the chemical network presented in Gong et al. (2017). This network (hereafter the GOW17 network) has been benchmarked against a highly accurate PDR code across a wide range of temperatures, pressures and metallicities (Gong et al. 2017). It accurately reproduces the CO abundances in a 1D equilibrium model, but in high density regions may over-produce C. Our implementation is a non-equilibrium, time-dependent 3D version of the above that contains several additional reactions that are unimportant in PDR conditions but that make the network more robust when dealing with hot, shocked gas. Full details of these modifications can be found in Hunter et al. (2021).

The self-shielding by CO of the non-ionizing UV interstellar radiation field (ISRF) and the $H_2$ self-shielding are modelled using the TreeCol algorithm developed by Clark et al. (2012) with a shielding length of 30 pc. The heating and cooling of the gas is computed simultaneously with its chemical evolution. We use the new molecular and atomic cooling functions from Clark et al. (2019). This improved treatment of gas cooling at high temperatures (T $\gg 10^4$ K) is key for modelling the effects of supernovae feedback in our simulations. Within the network we trace 9 individual non-equilibrium species: $H_2$, $H^+$, $C^+$, $CH_x$, $OH_x$, CO, $HCO^+$, $He^+$, $Si^+$ plus a further 8 which are derived from conservation laws or the assumption of chemical equilibrium: free electrons, H, $H_3^+$, C, O, $O^+$ He and Si. The pseudo-species $CH_x$ and $OH_x$ represent CH, $CH_2$, $CH^+$, $CH_2^+$ etc. in the first case, and OH, $OH^+$, $H_2O^+$, $H_2O$ etc. in the second case.

| Model name | Z($Z_\odot$) | Dust-to-Gas | $\zeta_H$ (s$^{-1}$) | $G_0$ (Habing Units) |
|---|---|---|---|---|
| Z.10 G.10 | 0.10 | 0.10 | $3.0 \times 10^{-18}$ | 0.17 |
| Z.10 G.01 | 0.10 | 0.10 | $3.0 \times 10^{-19}$ | 0.017 |
| Z.01 G.10 | 0.01 | 0.01 | $3.0 \times 10^{-18}$ | 0.17 |
| Z.01 G.01 | 0.01 | 0.01 | $3.0 \times 10^{-19}$ | 0.017 |

**Table 1.** Values used for metallicity ($Z$), dust-to-gas ratio (relative to the value in solar metallicity gas), cosmic ionisation rate ($\zeta_H$(s$^{-1}$)), and UV-field strength ($G_0$ (Habing Units)) in the initial conditions for each of the four simulations.

| | |
|---|---|
| $\rho_c$ (g cm$^{-3}$) | $2.4 \times 10^{-20}$ |
| n (cm$^{-3}$) | 1e4 |
| $r_{acc}$ (pc) | 1.75 |
| Softening Length (pc) | 2.0 |
| $\epsilon_{SF}$ | 0.1 |
| $r_{SNe}$ (pc) | 5 |

**Table 2.** Sink particle parameters used across all four simulations. $\rho_c$ is the sink density threshold, $r_{acc}$ is the accretion radius, $\epsilon_{SF}$ is the star formation efficiency and $r_{SNe}$ is the radius of scatter for SNe around the sink particle.

In our simulations we vary the metallicity with respect to the solar reference value by decreasing the carbon abundance, oxygen abundance, and dust-to-gas ratio in line with the reduction in Z. For simplicity we assume a linear scaling between the dust-to-gas ratio and Z, though in reality observations hint at a steeper relation at low Z (see e.g. Rémy-Ruyer et al. 2014). Similarly, we scale the magnitude of the cosmic ray ionisation rate of atomic hydrogen ($\zeta_H$(s$^{-1}$)) based on the magnitude of our chosen UV field strength ($G_0$), under the assumption that both scale linearly with the star formation rate. For our fiducial model we consider all variables to be 10% of solar ($Z = 0.10\,Z_\odot$, $G_0 = 0.17$). We then run 3 more simulations with the variables as seen in Table 1.

### 2.3 Sink Particles

For stars to form, gas within the ISM must undergo gravitational collapse to protostellar densities. Even at dwarf galaxy scales it is computationally unfeasible to simulate the entirety of the star formation process. In order to proceed we adopt the use of collisionless sink particles (sinks), a technique that has been widely adopted in computational studies of star formation (Bate et al. 1995; Federrath et al. 2010). Here we replace the densest collapsing and gravitationally-bound regions of a molecular cloud with a sink particle that can undergo future accretion. Our method is the same as that of Tress et al. (2020) who integrate sink particles in a galactic scale simulation successfully.

Sinks form when, within a given accretion radius, $r_{acc}$, a region with density greater than a defined threshold $\rho_c$ satisfies the following criteria:

(i) The gas flow must converge. To define this both the velocity divergence and the divergence of the acceleration must be negative.

(ii) The collapsing region must be centred on a local potential minimum.

(iii) The region cannot be within the accretion radius of any other sink particles, and should not move within the accretion radius in a time less that the local free-fall time.

(iv) The region must be gravitationally bound. To do this it must satisfy the conditions: $U > 2(E_k + E_{th})$, where $U = GM^2/r_{acc}$ is the gravitational energy of the region within the accretion radius, $E_k = 1/2\Sigma_i m_i \Delta v_i^2$ is the total kinetic energy of the gas within the accretion radius where $m_i$ and $\Delta v_i$ are the mass and velocity dispersion within cell $i$ and we sum over all cells within the accretion radius, and $E_{th} = \Sigma_i m_i e_{th,i}$ is the total internal energy of the region where $e_{th,i}$ is the specific thermal energy of cell $i$.

These ensure that a sink is only formed if the gas is truly self-gravitating and collapsing.

When a gas cell satisfies these conditions it is converted into a sink particle that is allowed to accrete mass. It can do this if the gas within $r_{acc}$ is above the density threshold and gravitationally bound to the sink particle. The amount of mass accreted onto the sink from the cell is calculated by:

$$\Delta m = (\rho_{cell} - \rho_c)V_{cell} \quad (1)$$

where $\rho_{cell}$ is the initial gas density of the cell and $V_{cell}$ is the volume. We limit the accreted mass to no more than 90% of the initial mass of the cell. Once the mass has been removed from the cell, the new density of the cell is the threshold density $\rho_c$, and any other values that depend on the mass are updated as necessary. If the gas cell lies within the accretion radii of multiple sinks, the accreted mass is given to the sink with which it is most strongly bound. We set a maximum stellar mass threshold of each sink to 200 $M_\odot$. Above this, accretion is turned off, instead allowing another sink to form. In this way our sinks always resemble small stellar sub-clusters, and any large star-forming region will contain multiple sinks.

The time stepping used by AREPO allows gas cells to be active on different time steps to the sink particles. Accretion only occurs when a gas cell and the sink particle it is bound to are both active. So that accretion is not missed from cells that spend a short time within $r_{acc}$, the time step on which the the sinks are evolved is kept equal to the shortest gas cell time step.

To ensure that collapse and any fragmentation that might occur is properly resolved we ensure that the local Jeans length is resolved by at least 8 resolution elements (Truelove et al. 1997; Federrath et al. 2011). However, the Jeans length in the coldest and densest regions of a molecular cloud can become prohibitively small at high densities. To overcome this, a density threshold for refinement is implemented, $\rho_{lim} = 2.4 \times 10^{-20}$ g cm$^{-3}$, which also defines the sink creation density, i.e. $\rho_c = \rho_{lim}$. At densities above $\rho_{lim}$, we no longer refine the grid, however any unaccreted gas at these densities is still able act hydrodynamically with its environment and may continue to increase in density. This strikes a balance between computational efficiency and high resolution within dense regions.

The gravitational softening length for the collisionless sink particles is 2 pc as in Hu et al. (2016) to allow a comparison (but with higher resolution in our gas cells). Our sinks should therefore be thought of as small sub-clusters of stars, rather than as individual stellar systems. Other studies use a more stochastic approach that creates stars of a given mass in the densest regions of the ISM given an efficiency per free fall time. Our approach ties the star formation directly to the cold dense gas in the ISM allowing us to causally connect the sink mass to the material that is gravitationally bound.

### 2.4 Feedback

By using the chemical model discussed above with the cooling and self-shielding properties of molecular gas we are able to form molecular clouds. However, such clouds have a finite lifetime. Galactic shear has been shown to be too inefficient a disruptive process to produce realistic cloud lifetimes and masses (Jeffreson & Kruijssen 2018), and so stellar feedback must be included.

The sinks are associated with a discrete stellar population based on the formalism of Sormani et al. (2017). Using an initial mass function (IMF) a set of discrete mass bins are populated via a Poisson distribution with a carefully chosen mean for each bin. This ensures that the mass distribution of stars within a sink particle follows the chosen IMF even if the sink is too small to fully sample the IMF. For each star that is more massive that $8M_\odot$ formed in this way, a supernovae explosion (SNe) is generated at the end of the star's lifetime, according to their mass in Table 25.6 of Maeder (2009).

As sink particles represent groups of stars, we do not assume the related supernovae occur directly at the location of the sink. To position them, we randomly sample a Gaussian distribution centred on the sink's location with a standard deviation $r_{\rm SNe}$ = 5 pc, based on an average cloud size of 10pc.

We assume a SF efficiency in the sink of $\epsilon_{\rm SF}$ = 0.1. Our choice of $\epsilon_{\rm SF}$ is motivated by the fact that even at our sink creation density ($\rho_c$ = 2.4 × $10^{-20}$ g cm$^{-3}$, $n$ = $10^4$ cm$^{-3}$) star formation is still inefficient (Krumholz & McKee 2005; Evans et al. 2009). It is also consistent with the observed range of star formation efficiencies in dense molecular gas (Jeffreson et al. 2021; Krumholz et al. 2012; Evans et al. 2014; Heyer et al. 2016). Our sink particle model does not allow us to directly trace the composition of the sink, and so we make the assumption based on our $\epsilon_{\rm SF}$ that 90% of the mass of the sink is in the form of gas which is gradually returned to the ISM through our feedback mechanisms explained below. The remainder of the mass in the sink is considered to be locked in stars.

After the last SNe tied to a sink particle has occurred, the sink has the same mass as the stellar component alone. When this happens it gets turned into a collisionless N-body star particle. However, some sinks will not accrete enough mass to form a massive star. This means that the gas in the sink will not be returned to the ISM from a SNe. To account for this, after a period of 10 Myr, if a sink has not formed a massive star then the star is turned into a star particle and the remaining gas mass is returned to the ISM by uniformly adding the gas mass to all gas cells surrounding the sink out to 100 pc.

The energy injection from the supernova is modelled from a modified version of an algorithm originally implemented in AREPO by Bubel (2015). We calculate the radius of the supernova remnant ($R_{\rm sr}$) at the end of its Sedov-Taylor phase assuming a SNe energy of $10^{51}$ erg and measure a local mean density of $\bar{n}$. This is then compared to the injection region ($R_{\rm inj}$), a sphere around the SNe that contains 32 Voronoi cells. If $R_{\rm sr}$ > $R_{\rm inj}$ then we inject $10^{51}$erg into the injection region as thermal energy and have the gas become fully ionised. When this is not the case and the SN remnant is unresolved, then the local density is considered too high for thermal injection, as it would radiate away too quickly to generate strong shocks or deposit the correct amount of kinetic energy into the surrounding ISM. In response to this the code directly injects the correct terminal momentum instead, without changing the temperature or ionisation state of these regions. The injection radius is kept small so that we minimise when we inject momentum into the ISM instead of thermal energy. This method for injecting SNe into simulations and has been previously used successfully by a number of other authors (Kimm & Cen 2014; Hopkins et al. 2014; Walch et al. 2015; Simpson et al. 2015; Kim & Ostriker 2017).

Due to the relatively low stellar mass of a dwarf galaxy and the low expected rate of type Ia SNe in such a small population we include only type II SNe. Naturally, there are other forms of feedback that are associated with star formation and dispersing GMCs, eg stellar winds and radiation from young stars and photoionisation. These are important as they affect the ISM before SNe feedback commences (Dale et al. 2014; Inutsuka et al. 2015; Gatto et al. 2017; Rahner et al. 2019; Chevance et al. 2020). However, to include all forms of feedback remains computationally difficult, and as such we leave this for future work. Excluding photoionisation will likely lead to an under-estimation of H$^+$ and over-estimation of H, and so we place a caveat on this. Jeffreson et al. (2021) include HII region feedback mechanisms in a Milky Way like galaxy and see no effect on star formation compared to runs with only SNe feedback but do note a change in the ISM and cloud characteristics.

|  | Mass (M$_\odot$) | Scale length (kpc) | $h_z$ (kpc) |
|---|---|---|---|
| DM Halo | 2.00 ×$10^{10}$ | 7.62 | - |
| Gas disc | 8.00 ×$10^7$ | 0.82 | 0.35 |

**Table 3.** Parameters for the different galactic components

### 2.5 Initial Conditions

We set up the dwarf galaxy as a stable disc in isolation. It consists of two initial components: a dark matter halo and a gaseous disc. Stars are not initially included but are created from sinks created during the initial burst before the galaxy reaches the steady-state period which we use for analysis. The steady-state period has stars present throughout its entirety. The size parameters are summarised in Table 3.

The dark matter halo uses a spheroidal Hernquist (1990) profile following:

$$\rho_{\rm sph}(r) = \frac{M_{\rm sph}}{2\pi} \frac{a}{r(r+a)^3} \quad (2)$$

where $r$ is the radius of the sphere, $a$ is the scale-length of the halo, and M$_{\rm sph}$ is the mass of the halo.

The gaseous disc component follows a double exponential density profile:

$$\rho_{\rm disc}(R, z) = \frac{M_{\rm disc}}{2\pi h_z H_R^2} {\rm sech}^2\left(\frac{z}{2h_z}\right) \exp\left(-\frac{R}{h_R}\right) \quad (3)$$

where $R$ is the disc radius and $z$ is its height, $h_z$ and $h_R$ are the scale-height (0.35 kpc) and scale-length (0.82 kpc) of the disc, respectively.

The initial conditions are generated using the method from Springel et al. (2005). The values have been chosen to be broadly comparable to Hu et al. (2016). The total baryonic mass is 8.00×$10^7$ M$_\odot$, slightly higher that Hu et al. (2016). Metallicity, the dust-to-gas ratio, cosmic ray ionisation rate and the UV-field are all set to 10% solar for the fiducial simulation. It is generally assumed that the relation between the star formation rate surface density and $G_0$ is linear (Ostriker et al. 2010), but to test this we vary the the above parameters between 1% and 10% of the solar neighbourhood value. Table 1 shows the values used for the four simulations. The initial temperature of the gas is set to $T$=$10^4$ K and the composition is initially fully atomic.

### 2.6 Resolution

Within the simulations we set a base mass resolution for the gas cells of 100 M$_\odot$, for the first 300 Myr, and then set it to 50 M$_\odot$ for the remainder of the simulation. The code will refine/derefine a cell to keep them within a factor of 2 of this mass. However, on top of this we add a more stringent requirement that the local Jeans length (JL) is resolved by at least 8 resolution elements. Due to the Jeans refinement, in the densest regions of the ISM we resolve to below a solar masses as shown in Figure 2. For a total gas number density of $n = 100$ cm$^{-3}$ we have a cell radius ($r_{eff}$) of 0.50 pc, and at $n = 10^4$ cm$^{-3}$ a $r_{eff}$ is 0.16 pc (Figure 2). These high spatial resolutions are essential for accurately capturing the fragmentation behaviour of the dense gas clouds where our sink particles will form (Truelove et al. 1997).

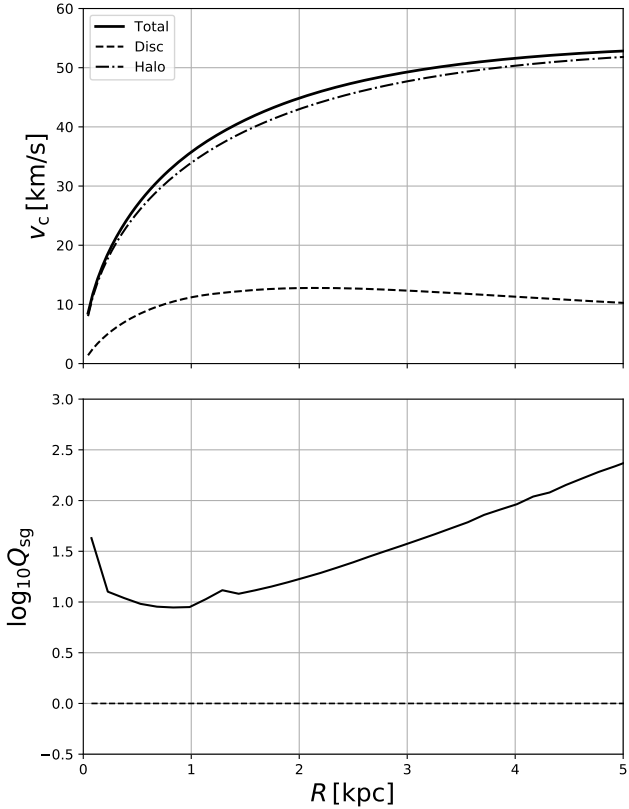

**Figure 1.** *Top*: The initial circular velocity curve. The solid line is the full dwarf galaxy, whilst the other lines show the contribution for the other components. *Bottom*: Combined star-gas Toomre parameter using the equation derived by Rafikov (2001)

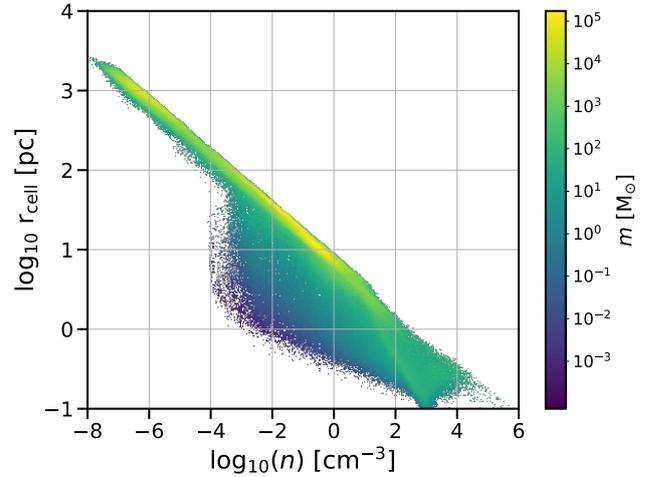

**Figure 2.** Cell radius as a function of number density. $r_{cell}$ is the radius of a sphere that has the same volume as the cell, taken from the Z.10 G.10 fiducial simulation at 500 Myr. Our refinement scheme devotes most of the computational power to regions of high density where the Jeans length criterion determines the cell mass. In these regions we reach sub-parsec resolution at densities that are comparable to the sink formation threshold of $10^4$ cm$^{-3}$.

As AREPO uses an adaptive mesh, the gas cells range in size and mass and so we cannot use a unique gravitational softening length. We therefore apply adaptive softening to the gas cells, with a softening length that varies with the cell radius. For the dark matter, we use a fixed softening length of 640 pc.

## 3 RESULTS

In this section we look at the resulting star formation of the dwarf galaxy models. We first investigate the morphology, general structure and evolution of the galaxies. We then look at the effect that varying the parameters has on the underlying chemistry, mass fractions and phases of the ISM. Finally we focus on the SFR and how the different metallicities and UV fields affect the KS relationship.

### 3.1 Morphology

We run the simulations for 1000 Myr, allowing a steady state to develop in all runs, which we find is achieved for the star formation rate after 300 Myr. In all four models we have an initial burst of star formation during the first 200 Myr. This arises due to the lack of supportive supernova feedback at the beginning of the simulation and leads to rapid collapse of the gas and sink particle formation. After a few Myr, the sinks evolve to the point where SNe can occur. This releases large amounts of thermal and kinetic energy into the disc in a short period of time. The supernovae push the gas out of the central regions, creating high density gas at the edge of a void within a few tens of Myr. All star formation happens within these dense regions at this time. Over time the gas falls back into the galaxy centres, until by 500 Myr there is little evidence of the initial burst in the morphology of the gas. We define the steady state period based solely on the star formation rate and not the overall morphology of the gas. As the burst is a consequence of our initial conditions, we focus our analysis of star formation on the steady state part of the evolution.

Visually the distribution of H$_2$ in Figure 3 differs substantially between the different models. As expected there is more H$_2$ in the 0.1 Z$_\odot$ models, than the 0.01 Z$_\odot$ models. In the G.10 models we see smaller regions of H$_2$, with it residing mostly in a thin layer in the disc. Whilst in the G.01 simulations there is a greater distribution of H$_2$; it is more dispersed and broadly follows the HI distribution. Looking through the plane of the galaxies (Figure 4) we can see that in the G.01 simulations, the H$_2$ is dispersed over greater scale heights.

Figure 5 shows the radius ($R_{95}$) within which 95% of the mass of H$_2$ (solid lines), sink particle mass (dotted lines), and HI (dashed lines) is contained. After 300 Myr a steady state forms. In all simulations the HI radii is relatively stable or smoothly varying, with the radius decreasing in the 0.1 Z$_\odot$ models to roughly the same radius as the 0.01 Z$_\odot$ models by 600 Myr. For H$_2$, $R_{95}$ is approximately constant throughout, with only very small scale variations due to the stochastic nature of the star formation within. In all cases the H$_2$ resides at substantially smaller radii than the HI, by at least a factor of two.

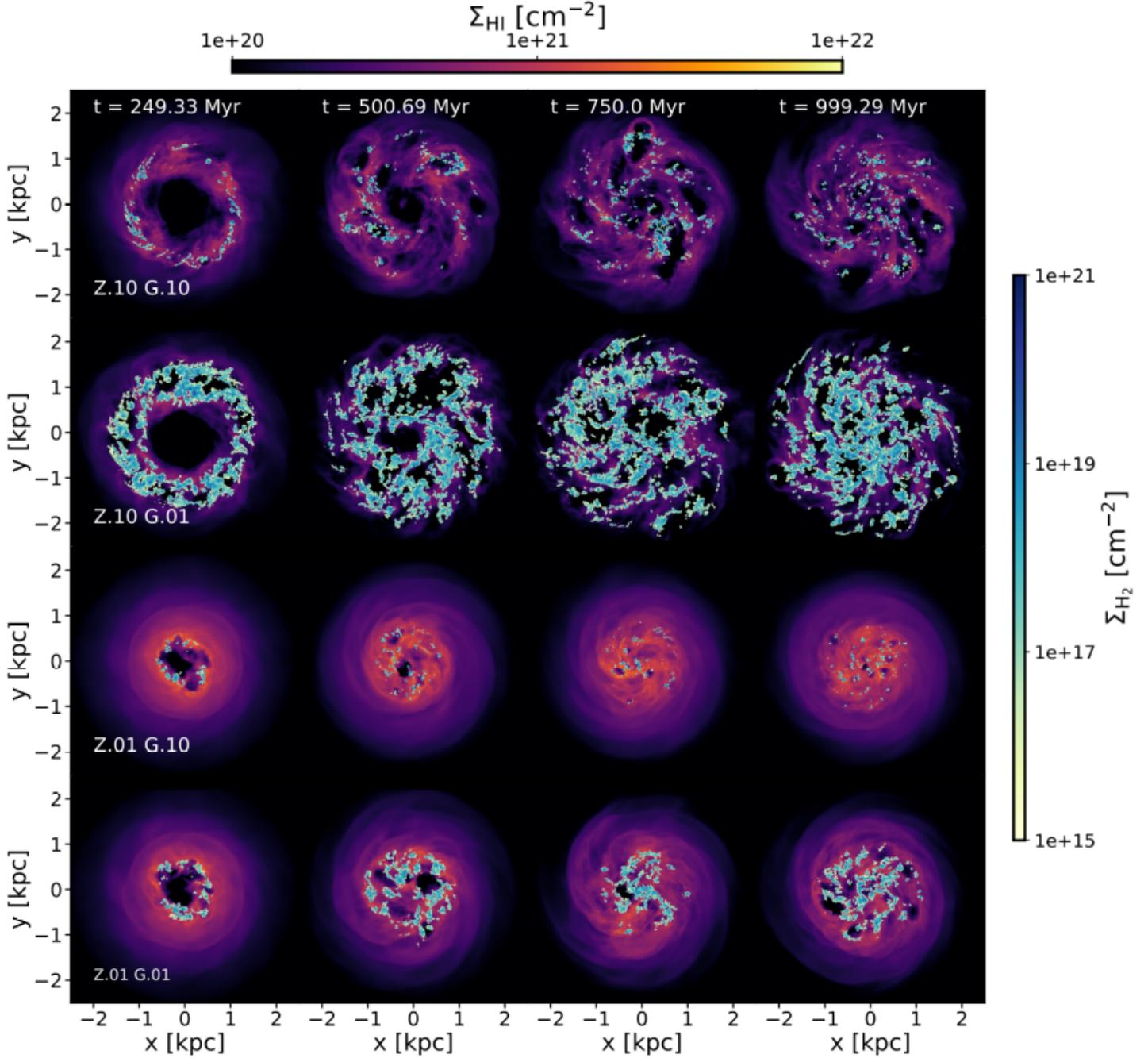

**Figure 3.** Plots of HI surface density ($\Sigma_{HI}$) and H$_2$ surface density ($\Sigma_{H_2}$) for the four simulations. From left to right, we show the disk at times $t$ = 250 Myr, 500 Myr, 750 Myr and 1000 Myr. From top to bottom, the rows correspond to runs Z.10 G.10, Z.10 G.01, Z.01 G.10 and Z.01 G.01. The H$_2$ surface density distribution varies significantly between the four different runs.

The metallicity is a greater determinant of the H$_2$ and sink particle $R_{95}$ radius than the UV field. In the $0.01Z_\odot$ models, the H$_2$ is confined to radii of 1 kpc or less. For the $0.1Z_\odot$ models, the H$_2$ lies at radii of 2 kpc or less. In this case the UV field plays a larger role, with the G.10 model lying slightly below the G.01 model (Table 4).

The most noticeable impact on the morphology is where the active star formation regions lie, as traced by the sink particles. Table 4 compares the $R_{95}$ radius of the gas phases discussed above, with the radius within which 95% of the sink mass is enclosed. In all cases, the $R_{95}$ for the sinks is very similar to that for the H$_2$ and much smaller than that for HI. In fact, the sink $R_{95}$ value is slightly larger than that of the H$_2$ due to sinks drifting away from the gas over time. We see from the table that star formation is confined to the the central $\sim 2$ kpc for the $0.1Z_\odot$ model, and $\sim 1$ kpc for the $0.01Z_\odot$ model.

|  | $R_{95}$ HI [kpc] | $R_{95}$ H$_2$ [kpc] | $R_{95}$ Sinks [kpc] |
|---|---|---|---|
| Z.10 G.10 | 4.03 | 1.50 | 1.65 |
| Z.10 G.01 | 4.30 | 1.81 | 2.00 |
| Z.01 G.10 | 3.76 | 0.83 | 0.92 |
| Z.01 G.01 | 3.74 | 0.93 | 1.07 |

**Table 4.** Average steady state radius in kpc within which 95% of the mass lies for HI, H$_2$ and sink particles for each of the four simulations. Metallicity has a greater impact than $G_0$ on the values of $R_{95}$ for the H$_2$ and the sink particles.

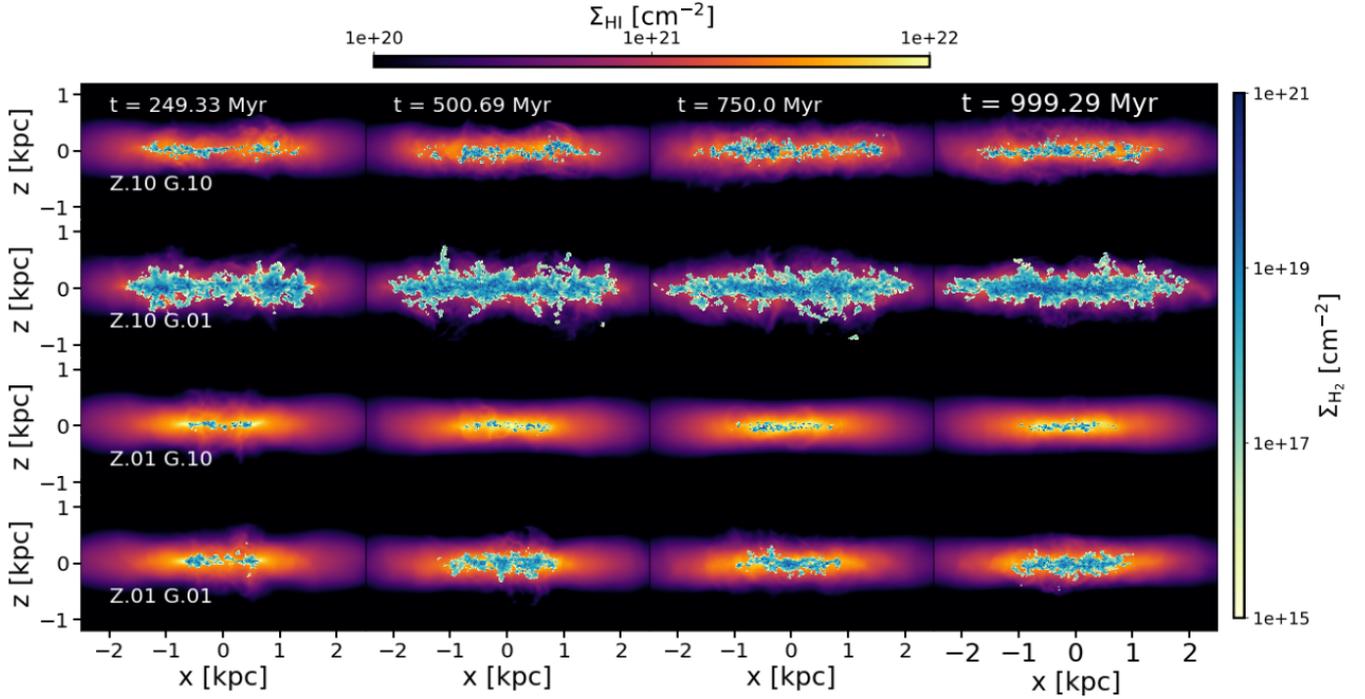

**Figure 4.** Plots of HI surface density ($\Sigma_{HI}$) and H$_2$ surface density ($\Sigma_{H_2}$) for the four simulations shown through the plane. From left to right, we show the disk at times $t = 250$ Myr, 500 Myr, 750 Myr and 1000 Myr. From top to bottom, the rows correspond to runs Z.10 G.10, Z.10 G.01, Z.01 G.10 and Z.01 G.01. We see that the scale height of the H$_2$ in the $z$ direction varies substantially from simulation to simulation but displays little variation over time.

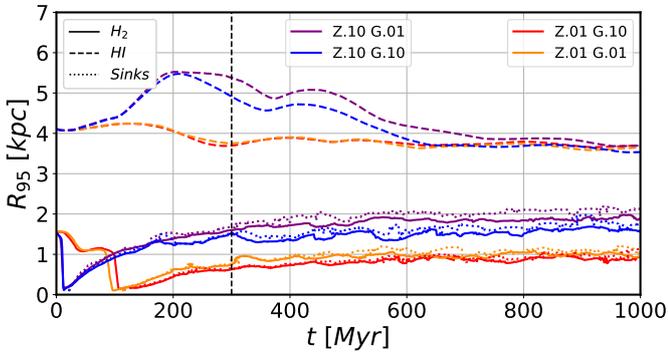

**Figure 5.** Radius within which 95% of the mass of H$_2$, HI and sinks lie over the duration of the simulations. The blue line is the Z.10 G.10 model, the purple the Z.10 G.01 model, the orange the Z.01 G.01 model and the red the Z.01 G.10 model. The vertical dashed line represents the time at which we consider the steady state to begin. The values of $R_{95}$ for the H$_2$ and the sink particles are very similar and are stable over time. Both are at least a factor of two smaller than the value of $R_{95}$ for the HI.

Figure 6 shows the average surface density radial profile for the HI, and H$_2$ over the steady state period. The HI density profile is largely unaffected by metallicity or UV field, though there is a small increase in the central 1 kpc in the $0.01Z_\odot$ models. The H$_2$ profile differs in line with Figures 3 and 5. For the $0.01Z_\odot$ models, the H$_2$ drops off rapidly, but for the $0.1Z_\odot$ models we see a more steady decline in surface density out to $\sim 2$ kpc. Overall, the $0.10Z_\odot$ models are $\sim 0.5$ kpc more extended in radius.

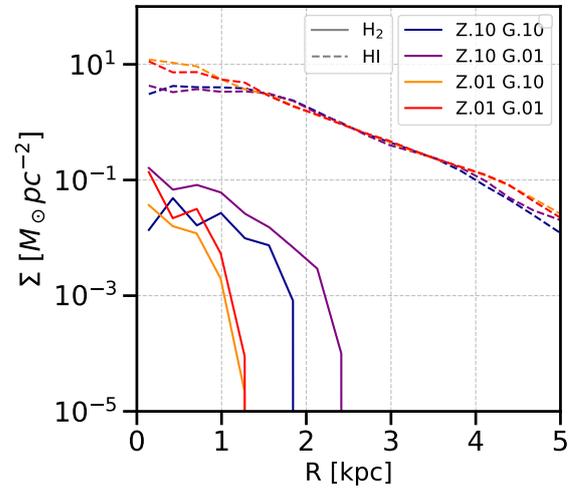

**Figure 6.** Average surface density radial profiles for the four simulations over the steady state period. The plot shows $\Sigma_{H_2}$, and $\Sigma_{HI}$. The $\Sigma_{HI}$ profile is more extended and has greater surface density than the $\Sigma_{H2}$ profile.

### 3.2 ISM Chemistry

In this section we present the main results of the evolution of the ISM and hydrogen species. We leave a more detailed analysis of the CO chemistry for future work since it is likely that the CO abundances are not fully converged at our current resolution (see e.g. Joshi et al. 2019).

Figure 7 shows how the $H_2$ mass fraction evolves with time in the four models in relation to total HI + $H_2$ gas mass. We note that the mass fraction of $H_2$ in all models is extremely small and the ISM is almost entirely dominated by HI. Table 5 shows the average steady-state mass fraction for $H_2$, the cold (T < 100K) HI + $H_2$ mass fraction and the $H_2$ ($\tau_{dep(H_2)}$) and cold ($\tau_{dep(cold)}$) depletion times, equation 4, in all four models. We also show the standard deviation in the depletion time ($\sigma_{\tau_{dep/ss}}$) as a guide to how much this varies within the disks. There is ∼ 7 − 8 times more $H_2$ in the $0.10Z_\odot$ models than in the $0.01Z_\odot$ models. These differences are most likely due to the reduced amount of dust surfaces on which to form $H_2$ in the lower metallicity case. Similarly, there is more mass in the cold phase in the $0.10Z_\odot$ models than in the the $0.01Z_\odot$ models. This is a consequence of the longer cooling times in the lower metallicity runs. Gas which has been shock heated or which is cooling from the warm neutral medium to the cold neutral medium following a compression takes much longer to reach a temperature < 100 K in these runs because there is little carbon or oxygen present to provide fine structure cooling. Therefore, even though the equilibrium temperature at each density varies only weakly with metallicity, the end result of reducing the metallicity is a reduction in the cold gas fraction. We also see that when we reduce the strength of the UV field by a factor of ten, we roughly double both the $H_2$ mass fraction (owing to the lower photodissocation rate) and the cold gas mass fraction (owing to the change in the photoelectric heating rate.

$$\tau_{dep(gas)} = \frac{M_{gas}}{SFR} \quad (4)$$

We note that as sink particles form in the coldest and densest regions of the ISM, they remove gas from these regions and lock it into collisionless particles. Unfortunately, we have no information on the density, temperature, or chemical composition of the trapped gas. Our sink particles have a formation radius of 1.75 pc and a proportion of the gas in this region may be molecular. Since we do not know the exact fraction that this corresponds to, we ignore any gas inside the sink in our main analysis. Thus, our results are a lower limit on the molecular gas. In Appendix B we consider how our results would change if the gas mass of young sinks (less than 3 Myr old) that have not had supernovae were included in our molecular total, in line with Olsen et al. (2021), a procedure which gives us an upper limit on the molecular gas mass. We do not include this contribution by default as it is unclear precisely how much of the gas locked up in the sinks is actually molecular.

We also examine the temperature phases of the ISM as a function of mass fraction of the total (HI + $H_2$) gas. Figure 8 shows two regimes, cold (below 100 K) and warm (100 − $10^{4.5}$ K). We do not consider the hot phase in this work as it is not relevant for the results discussed here. Like the $H_2$ mass fraction in Figure 7, the cold gas mass fraction shows small variations over the steady state period. Both plots are very similar, but the absolute values of the total cold gas fraction are higher than that of the $H_2$. Although there is a factor of 10 reduction in the metallicity between the Z.10 G.10 and Z.01 G.01 models, there is only a factor of 2-3 reduction in the amount of $H_2$ and cold gas. However it should be remembered that the warm gas and HI dominate in all simulations, with both the $H_2$ and cold gas having negligible mass fractions.

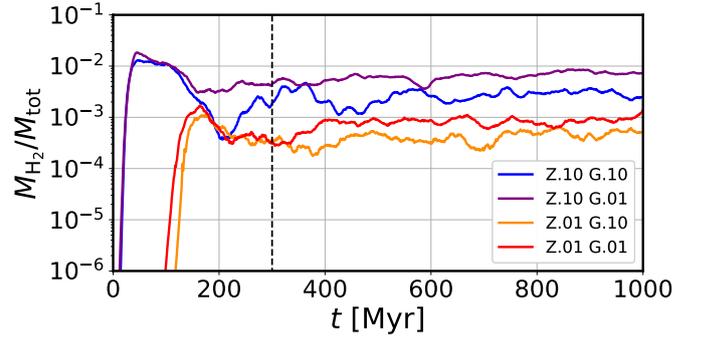

**Figure 7.** Mass fraction for $H_2$ for all 4 models. There is a higher mass fraction of $H_2$ in the $0.10Z_\odot$ models

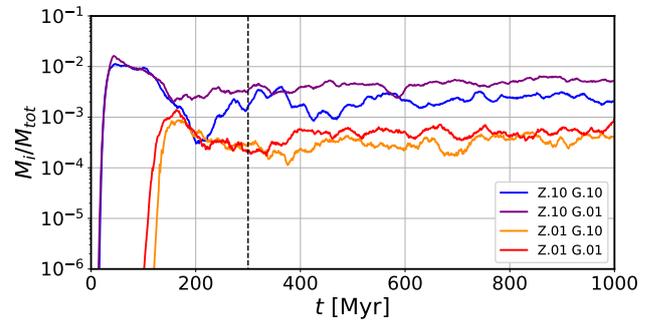

**Figure 8.** ISM temperature phase plot showing the mass fraction in the cold phase, which is considered to be below 100K. We do not include the warm phase as it is dominated by HI.

Figure 9 shows the $H_2$ mass-weighted phase diagrams (temperature vs number density). The majority of $H_2$ lies in the cold gas, with a large proportion of the mass below T = 100 K. In the low UV field models, we see a more diffuse distribution, with a significant fraction of the molecular gas in the warm phase, T = 100 − $10^{4.5}$ K, owing to the less effective photodissociation of $H_2$ in these runs. The $H_2$ distribution does not solely reside in the conditions typical of star forming regions (T < 100 K and $n$ > 100 $cm^{−3}$). This is especially true in the low UV field models, where non-star forming regions can contain a substantial fraction of the total amount of $H_2$ present in the disk.

Figure 10 shows the $H_2$ and cold gas depletion times for all four models. The $H_2$ depletion times for the $0.10Z_\odot$ models are over an order of magnitude faster than the values commonly observed in local spiral galaxies ($\tau_{dep}$ ∼ $2.0 \times 10^9$ yr) (Bigiel et al. 2008; Leroy et al. 2008), and those for the 0.01 $Z_\odot$ runs are smaller still. The cold gas depletion times are of less than an order of magnitude faster than that observed in solar metallicity spirals. However, as we will see later, this is not because star formation is more efficient in these systems. Rather, it is because the stars that are forming in these low metallicity systems are forming in cold HI-dominated gas, and so the $H_2$ depletion time gives a misleading view of how rapidly the gas is actually being consumed. Unlike in Tress et al. (2020), who use the same sink particle and feedback mechanisms, we do not see a decrease in $\tau_{dep}$ over time. In their simulation, this occurred because over the course of the simulation, the ongoing star formation appreciably depleted the amount of gas available for forming further stars. This does not happen in our simulations owing to the much smaller star formation rates involved, as we discuss further in the next section.

| | $M_{H_2}/$ $M_{HI+H_2}$ | $M_{cold}/$ $M_{HI+H_2}$ | $\tau_{H_2 dep/ss}$ (yr) | $\sigma_{\tau_{H_2 dep/ss}}$ | $\tau_{colddep/ss}$ (yr) | $\sigma_{\tau_{colddep/ss}}$ |
|---|---|---|---|---|---|---|
| Z.10 G.10 | 0.27% | 1.66% | 9.16 ×10$^7$ | 2.63 ×10$^7$ | 5.66 ×10$^8$ | 1.65 ×10$^8$ |
| Z.10 G.01 | 0.63% | 4.26% | 1.18 ×10$^8$ | 2.56 ×10$^7$ | 8.04 ×10$^8$ | 1.77 ×10$^8$ |
| Z.01 G.10 | 0.04% | 0.29% | 5.25 ×10$^7$ | 3.72 ×10$^7$ | 3.75 ×10$^8$ | 2.68 ×10$^8$ |
| Z.01 G.01 | 0.08% | 0.58% | 3.54 ×10$^7$ | 1.73 ×10$^7$ | 2.57 ×10$^8$ | 1.61 ×10$^8$ |

**Table 5.** Mass fractions and depletion times for $H_2$ and cold HI + $H_2$ gas. There is a smaller $H_2$ mass fraction and depletion time in the $0.01Z_0$ models. The cold gas depletion times are greater than the $H_2$ due to this being HI dominated.

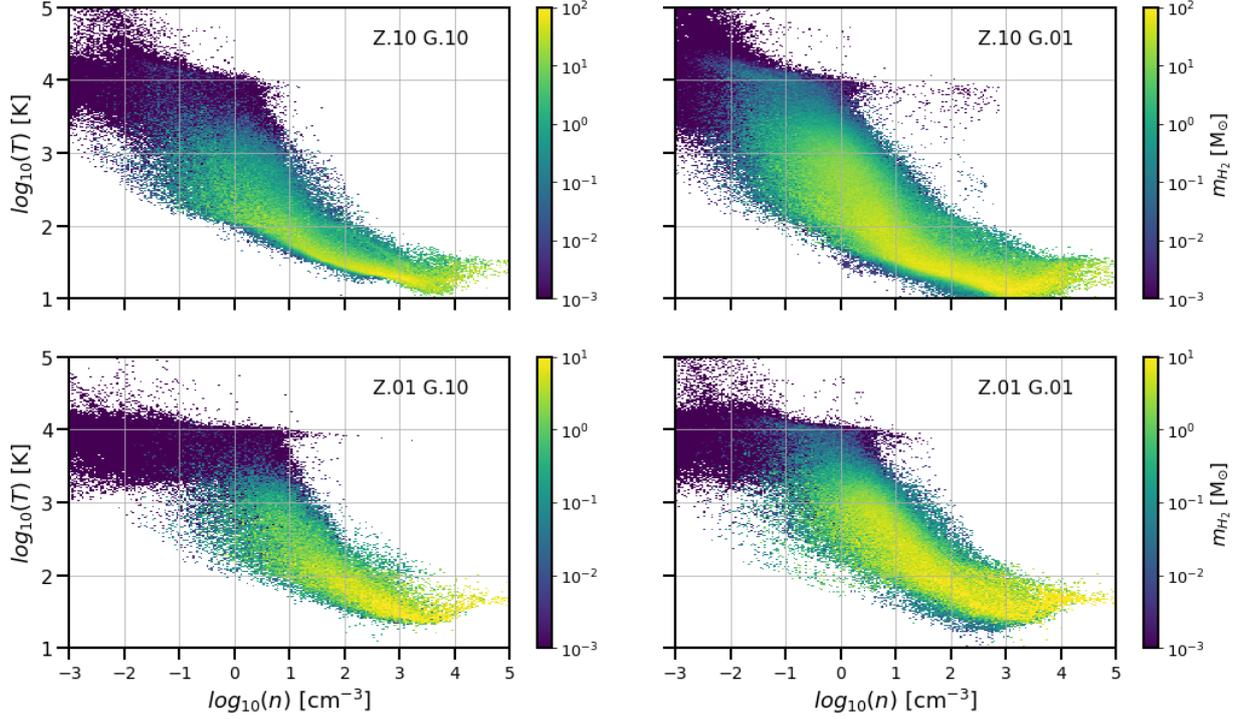

**Figure 9.** $H_2$ mass-weighted phase diagrams at $t = 500$ Myr. The majority of the $H_2$ mass lies in the cold (T < 100K) gas, but there is also a significant proportion of it in the warm diffuse phase in the G.01 models that arises due to the lower photodissociation rates in the in the very low $G_0$ simulations.

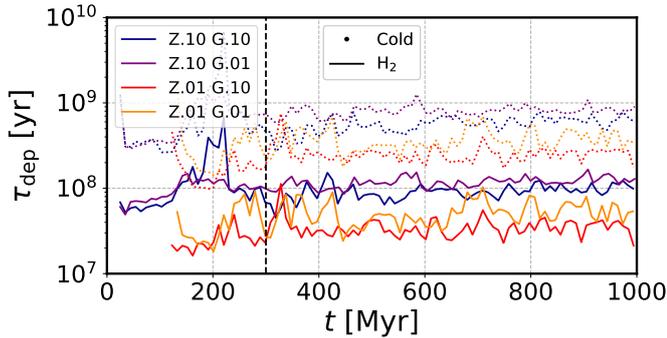

**Figure 10.** Depletion time ($\tau_{dep}$) for $H_2$ and cold gas as a function of time. $\tau_{dep}$ is shorter and more variable when we lower the metallicity of the models, but there is little change when the UV field strength is lowered.

### 3.3 Star Formation

In this section we present the results of our star formation model and how it is affected by the different conditions. For our star formation rate we use the stellar component of the sink particle mass ($\epsilon_{SF}$ × total sink mass) and convert it into a rate by taking the stellar mass accreted ($M_{\odot,acc}$) onto the sink particle at each snapshot plus any new sinks created ($M_{\odot,new}$) and dividing by the time difference between each the current snapshot ($t_2$) and the previous snapshot ($t_1$):

$$SFR = \frac{(M_{\odot,acc} + M_{\odot,new})}{t_2 - t_1} \quad (5)$$

In Figure 11 we show the steady state period for the star formation rate (SFR) as a function of time. The shaded regions show the $1\sigma$ deviation. The bursty nature of SF is clear in this plot.

Table 6 shows the average steady-state star formation rate (SFR$_{ss}$) and the standard deviation ($\sigma_{ss}$) for each model, along with the HI + $H_2$ surface density ($\Sigma_{HI+H_2}$) and $H_2$ mass fraction for comparison. There is little variation in the average SFR$_{ss}$ across the models. The key result is the similarity between the Z.10 G.10 and Z.01 G.01 models, where a factor of 10 reduction to both parameters has a negligible effect on the star formation rate, with both lying well within $\sigma_{ss}$ of each other. Comparing the SFR$_{ss}$ with the $\Sigma_{HI+H_2}$ and M$_{H_2}$/total values in Table 6, there is little connection to the total gas surface density, but there is a simple correlation to the H$_2$ mass fraction as when the SFR$_{ss}$ is increased, so is the mass fraction.

|  | SFR$_{ss}$ [M$_\odot$yr$^{-1}$] | $\sigma_{ss}$ | $\Sigma_{HI+H_2}$ [M$_\odot$pc$^{-2}$] | $\Sigma_{H_2}$ [M$_\odot$pc$^{-2}$] | M$_{H_2}$/ M$_{HI+H_2}$ |
|---|---|---|---|---|---|
| Z.10 G.10 | 2.01 ×10$^{-3}$ | 47.4% | 2.50 | 8.27 ×10$^{-3}$ | 2.7 ×10$^{-3}$ |
| Z.10 G.01 | 3.41 ×10$^{-3}$ | 26.7% | 2.40 | 10.67 ×10$^{-3}$ | 6.3 ×10$^{-3}$ |
| Z.01 G.10 | 7.65 ×10$^{-4}$ | 50.2% | 3.25 | 3.10 ×10$^{-3}$ | 0.4 ×10$^{-3}$ |
| Z.01 G.01 | 1.90 ×10$^{-3}$ | 42.8% | 3.11 | 5.78 ×10$^{-3}$ | 0.8 ×10$^{-3}$ |

**Table 6.** Average steady-state star formation rates, SFR$_{ss}$, the size of the 1$\sigma$ standard deviation as a percentage of SFR$_{ss}$, HI + H$_2$ surface density ($\Sigma_{HI+H_2}$), H$_2$ surface density ($\Sigma_{H_2}$) and H$_2$ mass fraction.

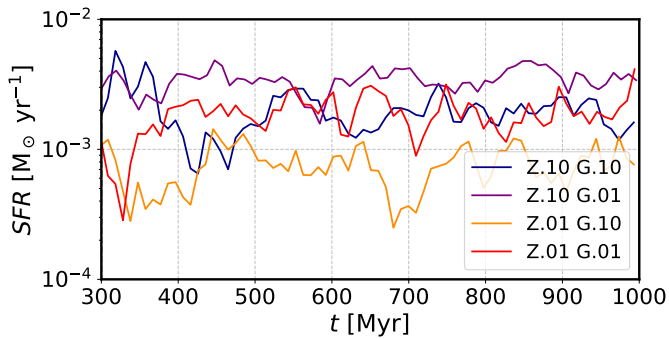

**Figure 11.** The star formation rate for the four models over the steady state period. The SFR of the Z.10 G.10 and Z.01 G.01 models is very similar.

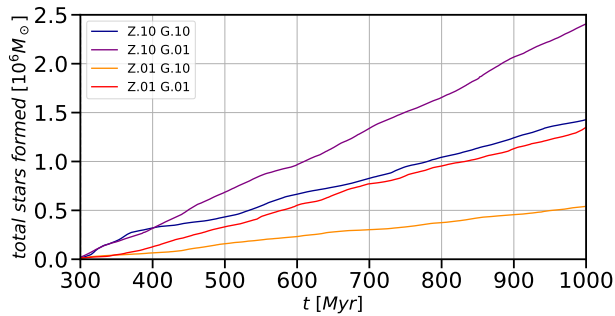

**Figure 12.** The cumulative mass of stars formed for the four models over the steady state period, which we take to start at $t = 300$ Myr. The similarity between the Z.10 G.10 and Z.01 G.01 models is clear.

The SFR$_{ss}$ for all four models are highly variable, or bursty over the steady state period. This is particularly true of the Z.01 G.10 model, where there are periods where the star formation rate falls well below $10^{-3}$ M$_\odot$ yr$^{-1}$.

Figure 12 shows the total mass of stars formed in each simulation over the steady state period. There is again a clear similarity between the Z.10 G.10 and Z.01 G.01 simulations, showing that the factor of 10 reduction in metallicity and UV field has had little effect on the total mass of stars formed. There is, however, a larger variation in the mass of stars formed when we vary the parameters separately, showing that the star formation rate is sensitive to a change in both the UV field and metallicity. In the two models where these quantities are varied independently, we see a factor of ∼ 5 difference in the SFR at 1 Gyr.

### 3.4 Kennicutt-Schmidt Relationship

In this section we examine the molecular gas Kennicutt-Schmidt (KS) relationship for each model. Figure 13 shows the KS relationship when averaging over a bin area of $(500pc)^2$.[1] The bins are evenly spaced boxes over the disc, within which the average gas column density and SFR are calculated. We stack data over the whole steady state period of the simulations, taking data every Myr, to ensure a complete sampling of the data; see Appendix A for an analysis of the time evolution. The colours represent the different models in the same format as all other figures. The grey dashed lines represent a linear relation of $N = 1$, as found by Bigiel et al. (2008) for a combination of spirals and late type/dwarf galaxies when considering H$_2$ surface density ($\Sigma_{H_2}$) with an aperture size of 400 pc, for a variety of different depletion times. For reference, the depletion time suggested by Bigiel et al. (2008) is approximately 2 Gyr. Our cold gas depletion time is almost an order of magnitude faster than the this, suggesting that the efficiency of forming stars in cold gas is similar in low metallicity dwarf galaxies and higher metallicity spirals, and that the low depletion times we see for the H$_2$ in the dwarfs are because the stars are forming in regions dominated by HI rather than because the H$_2$ is able to form stars more efficiently in these systems.

Table 7 shows the best fit power law indices and the 1$\sigma$ standard deviation. The power laws are fitted using the SCIPY linear regression module and the standard deviation is taken over the whole data-set using NUMPY. For our fiducial case (run Z.10 G.10) the power law slope for molecular hydrogen) is almost linear with $N_{H_2} = 1.09$. However as the metallicity and UV field strength are changed there are small departures from this. Lowering the field strength steepens the slope slightly to $N_{H_2} = 1.18$. When the metallicity is reduced, variations become more pronounced. In run Z.01 G.10, the slope becomes mildly *sub-linear*, whereas run Z.01 G.01 has the steepest slope.

Table 7 also shows the power law slopes that would be found if instead of the H$_2$ column density we considered cold gas ($T < 100$ K). The values of $N_{COLD}$ are always super-linear and are consistently higher than $N_{H_2}$. Thus there is proportionally more cold gas at higher column densities per unit star formation than there is molecular hydrogen.

We do not fit a slope to $\Sigma_{total}$ due to the limited column density range over which most of the data lies meaning that any fit is likely to be determined purely by the large scatter in our data and will therefore be inaccurate.

---

[1] In Appendix A we investigate the effects a larger bin of $(1kpc)^2$ would have upon our results as an analogy to using different telescope beam widths.

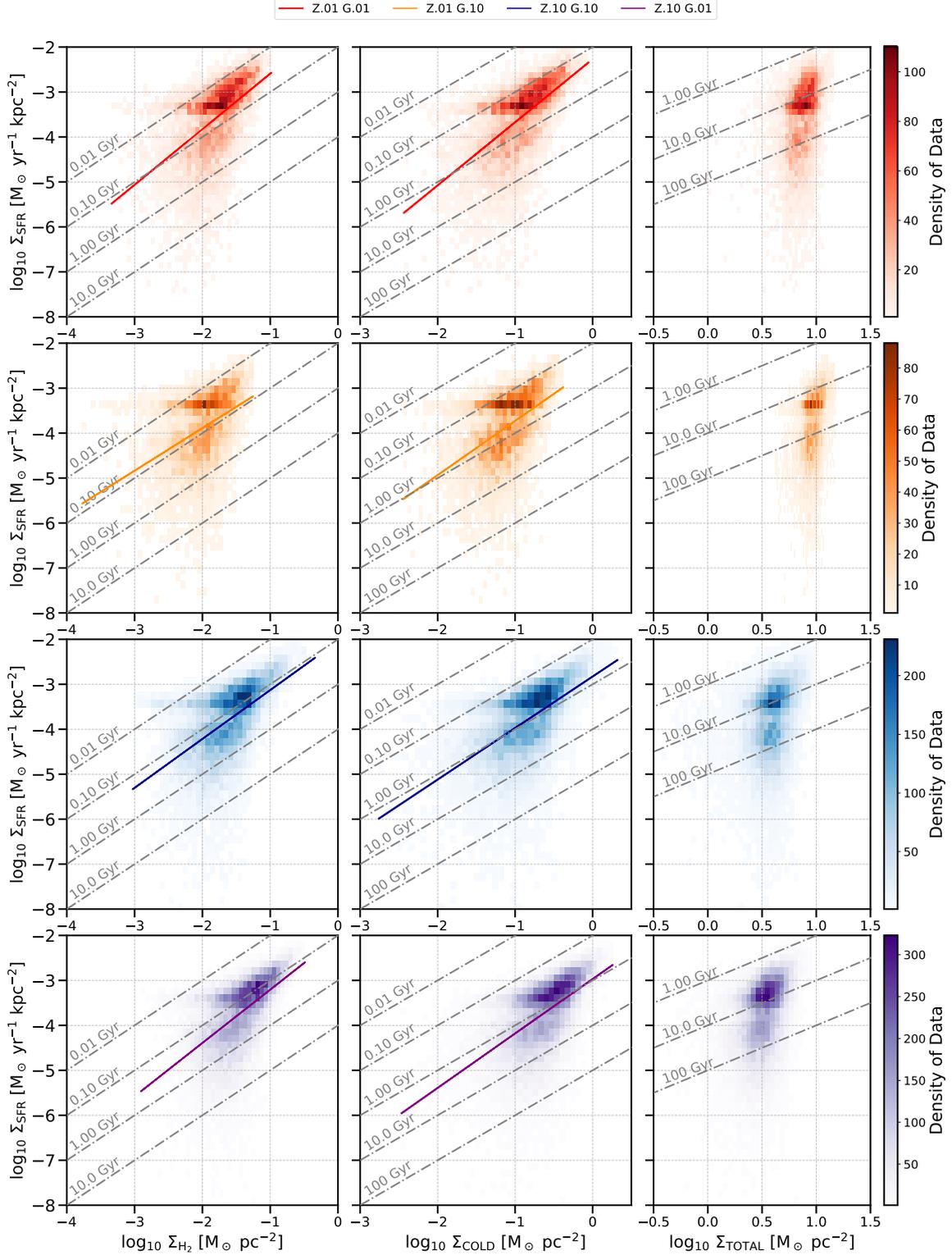

**Figure 13.** The Kennicutt-Schmidt relationship for the steady state time period for each model, averaging over a bin area of (500pc)$^2$. From left to right we show $N_{H_2}$, $N_{COLD}$ (where we define the cold phase to be all gas with $T < 100$ K), and $N_{total}$, where the total surface density includes all of the H and H$_2$, regardless of density, but does not include ionized gas. It is plotted as a heat map with the solid lines show the best-fitting power law for each simulation, and the grey dashed line represents the $N = 1$ slope for different depletion times. Results for different simulations are colour-coded as indicated in the key. We see that $N_{H_2}$ and $N_{COLD}$ both show linear to super-linear behaviour.

|          | $N_{H_2}$ (500 pc) | $N_{COLD}$ |
|----------|--------------------|------------|
| Z.10 G.10 | $1.09 \pm 0.014$  | $1.15 \pm 0.014$ |
| Z.10 G.01 | $1.18 \pm 0.012$  | $1.21 \pm 0.012$ |
| Z.01 G.10 | $0.95 \pm 0.027$  | $1.20 \pm 0.028$ |
| Z.01 G.01 | $1.24 \pm 0.022$  | $1.40 \pm 0.021$ |

**Table 7.** Power law slopes ($N$) for the Kennicutt-Schmidt law for the four models based on $\Sigma_{H_2}$ and $\Sigma_{COLD}$ with the standard errors from linear regression statistics. The molecular and cold KS slope is linear to super-linear, steepening as the UV field strength is reduced.

## 4 DISCUSSION

We have run four hydrodynamical models of metal poor dwarf galaxies varying both the metallicity and the UV-fields. They were run for 1 Gyr, and analysed over a steady state period that began after 300 Myr. We present here a discussion on the key points in this paper.

### 4.1 Morphology

Comparing the morphology of our simulations to those of Hu et al. (2016, 2017), we note little difference in the models that have similar conditions. Across our four models, the surface density radial profiles and scale heights are similar to their works. In agreement with their work, we find that the SFR and the general properties of the gas distribution reach a steady state after around 300 Myr.

All four models are dominated by HI, which shows high density structures and SN-driven voids (see Figure 3). Before the steady state period begins, we can see the $H_2$ traces the dense HI regions formed by the initial starburst. Figure 14 shows that $H_2$ does not always trace the densest regions of HI. For example, there is a dense $H_2$ cloud that appears to be in a void, though the mean column density in this void is around $10^{19}$ cm$^{-2}$. This occurs due to the HI being more easily disrupted by SNe.

Star-forming regions (represented by sink particles in our simulations) can drift apart from the both the HI and $H_2$ gas as shown by their greater radial extent in Table 4. Some of this is due to SNe occurring around the sink particle driving away the HI and creating the voids. Our sink particles are purely gravitational objects but the gas experiences additional hydrodynamic forces, which can also cause the two distributions to decouple. Figure 14 shows the location of sink particles in red, compared to the $H_2$ in blue. For the most part, the sinks lie at the centre of $H_2$ clouds, but there are instances where the sink is at the edge of a molecular region, or entirely separate from them. This is most likely due to the decoupling effect. However, it is theoretically possible for a sink to form in purely atomic gas as our sink formation criteria depend on the boundedness of the gas, and not on its molecular content. We will investigate the local environment of star formation within our dwarf galaxies in more detail in future work.

### 4.2 ISM Chemistry

The key points we take away from our treatment of the ISM and varying the metallicity and UV field strength are that a factor of 10 reduction in both variables has a large impact on the $H_2$ formed in the galaxy. As we decrease $Z$ or increase the field strength, the molecular gas becomes more closely confined to the galaxy centre (as seen by the shrinking of $R_{95}$).

In line with previous work, we find that the ISM in our simulated galaxies is dominated by HI, the majority of which is in the warm phase. The cold gas fraction ranges between 0.3% and ~ 4%, and the $H_2$ fraction is around 6-7 times smaller still. As expected, both the cold gas and the molecular gas fractions are highest in the Z.10 G.01 model, as it combines the highest metallicity and dust-to-gas ratio of our models with the lowest photodissociation rate.

|          | Cold $H_2$ | Warm $H_2$ |
|----------|------------|------------|
| Z.10 G.10 | 83%       | 17%        |
| Z.10 G.01 | 73%       | 27%        |
| Z.01 G.10 | 77%       | 22%        |
| Z.01 G.01 | 60%       | 39%        |

**Table 8.** Fraction of the total amount of $H_2$ in the cold phase ($T < 100$ K) and in the warm phase ($100 < T < 10^{4.5}$ K). The amount of $H_2$ in gas hotter than $10^{4.5}$ K is negligible. We see that a significant fraction of the $H_2$ content of each galaxies lies in the warm phase, particularly in run Z.01 G.01. Nevertheless, the majority of the $H_2$ is found in the cold phase in each case.

Figure 8 shows the phase space distribution in terms of the density and temperature of the molecular gas. While the bulk of the $H_2$ lies at temperatures below 100K, and would therefore be considered cold gas in our analysis, there is still a noticeable component at higher temperatures. This is particularly true in the simulations with lower UV field strengths, where there is a larger warm diffuse component to the $H_2$. For example, in the Z.01 G.01 model, 39% of the $H_2$ lies in the warm phase and 61% in the cold phase (Table 8).

In contrast to many previous studies, our sink particle based prescription for star formation does not prescribe a depletion time in cold and/or molecular gas, allowing us to explore how this varies in our simulated galaxies as a function of metallicity and UV field strength. Looking at Table 5 and Figure 10, we see that the molecular gas depletion time decreases significantly with decreasing metallicity. However, this behaviour merely reflects the fact that the dense clouds in which the stars form are dominated by atomic gas. Decreasing the metallicity decreases the molecular content of the clouds (because of the reduction in the dust content of the gas and hence the $H_2$ formation rate) but does not strongly affect the rate at which these clouds form stars, as we discuss in more detail below. This can be seen quite clearly if we look at the cold gas depletion times, which range from $\sim 3 \times 10^8$ yr to $10^9$ yr. These values are still somewhat smaller than the value of $2 \times 10^9$ years typical of molecular gas in late type spiral galaxies (Bigiel et al. 2008), but are an order of magnitude larger than the $H_2$ depletion time.

It is also interesting to compare the total gas depletion times from all models (Figure 15) with the total gas depletion times measured in the HI-dominated dwarf galaxies in the FIGGS survey (Roychowdhury et al. 2015). In our simulations, we typically find values between $10^{10}$ yr and a few times $10^{10}$ yr, in good agreement with the FIGGS results for galaxies with a similar mean gas surfac density (see e.g. Figures 4 and 5 in Roychowdhury et al. 2015).

### 4.3 Star Formation

Despite the large reduction in the $H_2$ mass, *the star formation rate is essentially unchanged* between the Z.10 G.10 and Z.01 G.01 models. This shows that the relationship between $H_2$ and star formation breaks down at low metallicity and low UV field strength. This is in line with the work of Glover & Clark (2012b); Hu et al. (2016). We also note an agreement with Krumholz (2012), who predicted that at low metallicities, star formation should become increasingly associated with cold atomic gas. A greater effect upon the star formation is seen when the metallicity and UV field are varied independently of each other. Reducing the UV field, for a constant metallicity, raises the star formation rate, as the reduced photoelectric heating rate makes it easier to form cold, dense clouds in the ISM. On the other hand, reducing the metallicity, for a constant UV field, lowers the star formation rate, since the increase in the cooling time makes it harder to form cold clouds. When both variables are varied their impacts cancel out, to a large extent.

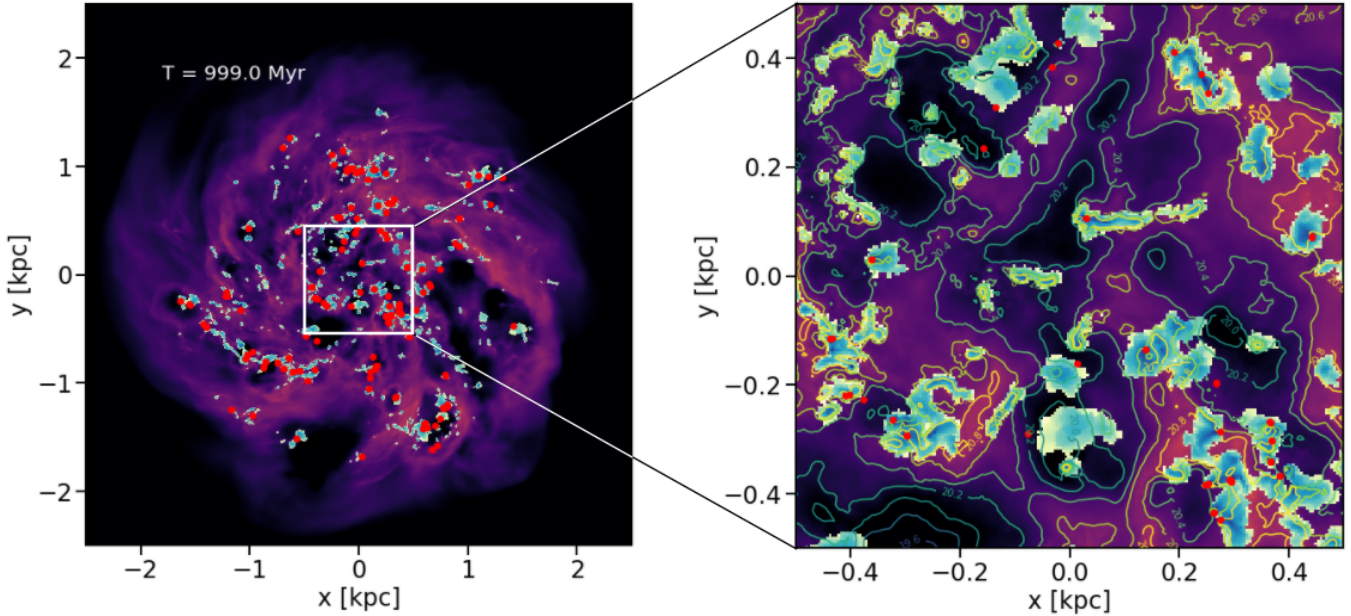

**Figure 14.** A close up of the central 0.5 kpc region of simulation Z.10 G.10 with sink particles shown as red dots. It shows some sink particles are situated in the voids, while some remain in the dense gas.

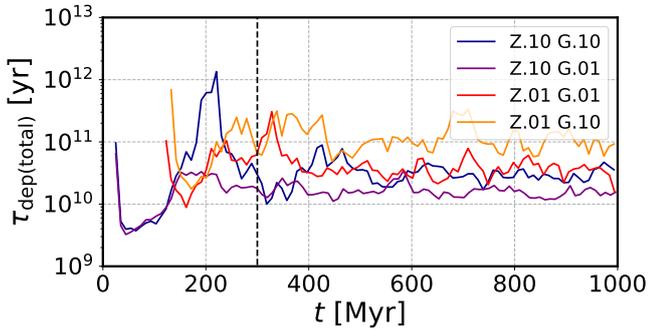

**Figure 15.** Depletion time for the total gas.

An obvious question then arises: what is the most realistic combination of variables in the extremely metal poor case? The UV field should in reality be generated by the stars. Since the star formation rate surface density we find in our simulated galaxies is roughly an order of magnitude smaller than that found in the disks of typical metal-rich spirals (Bigiel et al. 2008), we would therefore expect the UV field strength to be approximately an order of magnitude smaller as well. Therefore, we expect that the true UV field should on average look more like the value used in our G.10 runs than in our G.01 runs.

In our study, we keep the cosmic-ray ionisation rate selected for each simulation fixed as a function of time and position and do not include cosmic-ray injection from SNe. A recent study by

Dashyan & Dubois (2020) shows how SNe-driven cosmic-rays can affect the SFR in a large MHD dwarf galaxy simulation where they note a factor of $\sim 2$ reduction in the SFR when this is included. This likely means our $SFR_{ss}$ are an over estimation, but are still consistent with other comparable simulations. Jeffreson et al. (2021) show that by adding in HII momentum feedback in a Milky Way like galaxy at varying resolutions there is no effect on the SFR. This allows us to be confident that our star formation rates and models would be unaffected by additional feedback mechanisms other than SNe driven cosmic-rays.

Table 9 shows a comparison between our simulated $SFR_{ss}$ and surveys of Local Group dwarf galaxies. It is clear there is great variability in $Z_\odot$ and SFR, making a direct comparison difficult. For the Small Magellanic Cloud (SMC), with a mass of $4.20 \times 10^8$ $M_\odot$ and $Z = 0.1 Z_\odot$, the literature shows the SMC to have two star formation epochs: one 6 Gyrs ago, during which the average SFR was approximately 0.3 $M_\odot yr^{-1}$ and a more recent one at 0.7 Gyrs ago with a similar value for the SFR (Rezaeikh et al. 2014). During these bursts of star formation, the SMC has an SFR two orders of magnitude higher than we see in our simulations. Rezaeikh et al. and others note that this high SFR is likely to have arisen because of interactions with the LMC and the MW and note an overall SFR of $\sim 0.1 M_\odot yr^{-1}$.

We can also compare the SFRs in our simulated galaxies to those measured in NGC 147 and NGC 185, two satellite galaxies of Andromeda with similar masses to the SMC, but lower metallicities (Hamedani Golshan et al. 2017). One thing of note in their star formation history is an initial burst in SF early in their life which then becomes steady later on, similar to our simulations. Their SFRs are $\sim 2 \times 10^{-3} M_\odot yr^{-1}$ for NGC 147 and $\sim 5 \times 10^{-3} M_\odot yr^{-1}$ for NGC 185. These values are comparable to the $SFR_{ss}$ rates we see in our models.

It is clear that there is little-to-no direct connection between metallicity, SFR or mass on this small sample. No numerical scaling between the mass and SFR is noted, though it could be roughly assumed that more mass equates more star formation. NGC 147, however, doesn't follow this trend as it is twice as massive as our models but shows the same SFR. We also see that the specific star formation rate across our simulations and observations are comparatively similar.

### 4.4 KS Relationship

The customised version of AREPO used here, with time-dependent chemistry and the identification of bound collapsing regions via sink particles, allows us to produce a KS relationship that arises solely due to the star formation process with few assumptions.

Our results are in line with others in the literature (de los Reyes & Kennicutt 2019; Kennicutt & Evans 2012; Bigiel et al. 2008) in showing a linear to super-linear KS relationship with respect to both molecular hydrogen and cold gas. The slope becomes steeper for lower metallicity or lower UV field strength. While we use a linear regression analysis to match observational data, it should be noted that Shetty et al. (2014) applied a Bayesian analysis to a sample of nearby disk galaxies from the STING survey (Rahman et al. 2012) and found a sub-linear slope of N = 0.76 for $\Sigma_{H_2}$.

A one-to-one correlation between molecular gas surface density and SFR has been interpreted as showing that $H_2$ is a necessary precursor of star formation (Bigiel et al. 2008; Pessa et al. 2021). However, this is clearly not the case in our simulations. Although we recover a close to linear relationship between SFR surface density and $H_2$ surface density in each of our simulations, the normalization of this relationship – the $H_2$ depletion time – varies substantially as we change the metallicity. This is consistent with the behaviour that we would expect if the true correlation is between cold gas surface density and SFR surface density. In this picture, the correlation with $H_2$ arises because the surface density of $H_2$ separately correlates with the cold gas surface density (i.e. $\Sigma_{H_2} \propto \Sigma_{COLD}$), with a coefficient of proportionality – the mean $H_2$ mass fraction in the cold clouds – that depends on the metallicity and UV field strength.

It is also interesting to explore where deviations from a linear relationship between star formation and cold gas or $H_2$ arise. To investigate this, we show in Figures 16 and 17 box plots of the local depletion times in 500 pc bins over the steady state period used to calculate the KS relationships for different column density regimes. In both cases the depletion times are shorter at lower column densities where it is harder for both $H_2$ and cold gas to form. Thus the column density is lower for a given star formation rate, and the slope becomes shallower. This appears to be the origin of the sub-linear relation found in run Z.01 G.10, where the ambient UV field is particularly effective at dissociating $H_2$ and heating the gas.

At high column densities there is also a reduction in the depletion time, which is likely due to gas being more rapidly consumed by more efficient star formation due to the reduced free fall time at higher densities. This has the effect of steepening the slope, and is more pronounced at lower metallicities where the $H_2$ and cold gas reservoirs are smaller.

Observationally, it is hard to measure the $\Sigma_{H_2}$–$\Sigma_{SFR}$ KS relationship in dwarf galaxies as $H_2$ is very difficult to directly detect and other species such as CO are extremely faint. However, the $\Sigma_{HI}$–$\Sigma_{SFR}$ KS relationship has been studied by (Roychowdhury et al. 2015; Filho et al. 2016) and found to be super-linear. Roychowdhury et al. (2015) find a power law slope of 1.5 looking over different spatial scales for both spiral and dwarf galaxies, which is similar to other works (e.g. Kennicutt & Evans 2012). In addition, Filho et al. (2016) note that XMPs generally follow the literature but there is a fraction that fall off the slope that the authors suggest could be due to CO-dark H$_2$. de los Reyes & Kennicutt (2019) use various different fits to look at the $\Sigma_{HI}$-$\Sigma_{SFR}$ KS relationship. Each method gives a different slope, but they find their preferred fitting method gives a value of 1.26 ± 0.08 for a combination of spirals and dwarf galaxies. Comparing this to our results for H$_2$, they are shallower, but lie within one standard deviation of the de los Reyes & Kennicutt (2019) result.

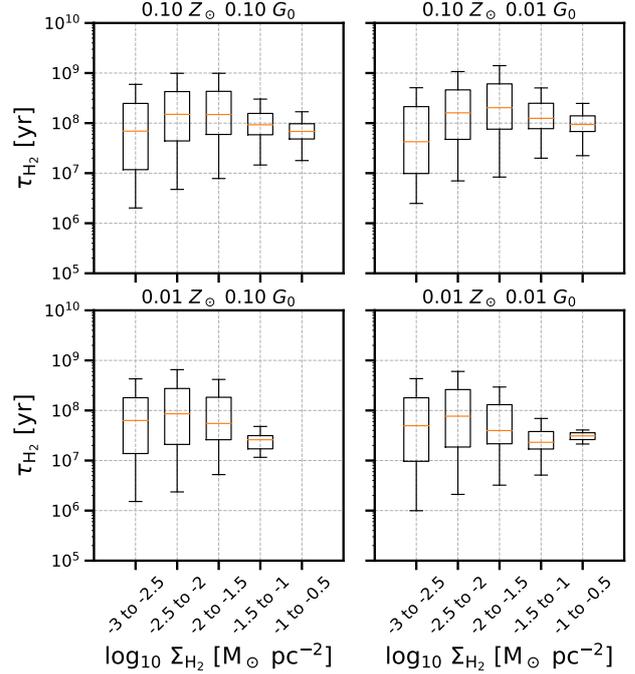

**Figure 16.** Local 500 pc scale $H_2$ depletion times binned in relation to $H_2$ column density. In all models we see a short depletion time in the most diffuse and densest $H_2$ with the longest times at intermediate column densities.

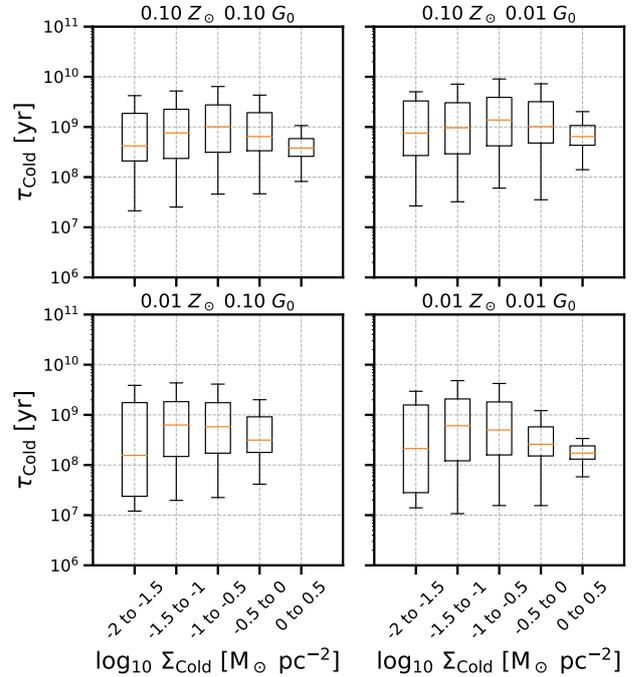

**Figure 17.** As Figure 16, but for the depletion time and surface density of the cold gas. In all models we see a a similar trend to the $H_2$ gas with shorter depletion time in the most diffuse and densest cold gas. The values are fairly consistent across column densities in the $0.10Z_\odot$ models but show more variability in the $0.01Z_\odot$ models.

|  | Mass ($M_\odot$) | SFR$_{ss}$ (M$_\odot yr^{-1}$) | $Z_\odot$ | specific SFR$_{ss}$ ($yr^{-1}$) |
|---|---|---|---|---|
| Z.10 G.10 | $8.0 \times 10^7$ | $2.01 \times 10^{-3}$ | 0.10 | $2.5 \times 10^{-11}$ |
| Z.10 G.01 | $8.0 \times 10^7$ | $3.41 \times 10^{-3}$ | 0.01 | $4.3 \times 10^{-11}$ |
| Z.01 G.10 | $8.0 \times 10^7$ | $7.65 \times 10^{-3}$ | 0.10 | $9.7 \times 10^{-11}$ |
| Z.01 G.01 | $8.0 \times 10^7$ | $1.90 \times 10^{-3}$ | 0.01 | $2.4 \times 10^{-11}$ |
| Milky Way | $9.5 \times 10^{10}$ | 8.25 | 1 | $8.7 \times 10^{-12}$ |
| LMC | $2.7 \times 10^9$ | 0.2 | 0.4 | $7.4 \times 10^{-11}$ |
| SMC | $4.2 \times 10^8$ | 0.1 | 0.1 | $2.4 \times 10^{-10}$ |
| NCG147 | $1.6 \times 10^8$ | $2 \times 10^{-3}$ | 0.1 | $1.3 \times 10^{-11}$ |
| NGC185 | $2.4 \times 10^8$ | $5 \times 10^{-3}$ | 0.08 | $2.1 \times 10^{-11}$ |

**Table 9.** Comparing the steady state SFRs, metallicities, masses and specific SFR of our models against some observed values. We use the value from our fiducial run for the SFR scaling ($2 \times 10^{-3}$). The values for the Milky Way come from Kennicutt & Evans (2012) for the gas mass and star formation rate and Ness & Freeman (2016) for an estimate on the global metallicity of the Milky Way. The LMC values are from Harris & Zaritsky (2009), the SMC ones from Harris & Zaritsky (2004), the SFRs for NGC 147 and NGC 185 come from Hamedani Golshan et al. (2017) and estimates of their metallicites are taken from Crnojević et al. (2014). The specific SFRs are calculated by dividing the SFR by the total mass of the galaxies as presented in column one of this table. There are all very similar with little variation across the models and different galaxies.

### 4.5 Caveats

The ISM is a complex system that has many active processes occurring within it. Any simulation of this system is inevitably a simplification and ours are no exception. In particular, in our simulations we do not currently include MHD, photoionisation, or the effects of stellar winds (although the effects of the latter are likely small at these metallicities).

As stated in Section 2.3, we do not form individual stars. Even though simulating a dwarf galaxy allows us to achieve extremely high resolution, we are still limited as to how fine we can go. As a result we cannot state how the ISM is behaving within the regions represented by the sink particles. In the bulk of this work we have excluded the sink particles when studying the gas properties, but in Appendix B we show how the H$_2$ depletion time and molecular KS relationship change if we assume that all of the non-star-forming gas in the sinks is molecular.

Another simplification made here is our assumption that the UV field is constant throughout the galaxy. This is unrealistic: in reality, it will be stronger closer to regions of active star formation and weaker away from these regions. Similarly, we neglect variations in the metallicity and dust-to-gas ratio within the disk, which is a reasonable description of some dwarf galaxies but not all (see e.g. the discussion in Section 5.1 of James et al. 2020).

Finally, we note that as we do not include photoionisation in our simulations, we likely over-produce diffuse atomic gas and under-produce diffuse ionised gas, particularly at large scale heights above the disk. However, we know from observations that there is far more atomic gas than ionised gas in most dwarf galaxies, and so we do not expect this limitation to significantly impact our results.

## 5 CONCLUSIONS

We have modelled four isolated dwarf galaxies for 1 Gyr using the AREPO code. In each simulation we varied the metallicity and/or the UV field strength. Our models include gas self-gravity, a time-dependent, non-equilibrium chemical network, local shielding, accreting sink particles to model star formation, and supernovae tied to the sinks. Our resolution reaches down to sub-parsec scales and resolves the Jeans mass throughout the simulation with a minimum of 8 resolution elements. The main results can be summarised as follows:

• The disks of our four simulated galaxies are dominated by warm atomic hydrogen. The cold gas mass fraction ranges from < 1% to a few percent, while the H$_2$ mass fraction is less than 1% in every case. Reducing the UV field strength leads to the formation of more H$_2$ and more cold gas; conversely, reducing the metallicity decreases the amount of both.

• H$_2$ is found in the warm phase in all models, particularly the ones with the lower UV field strength. This supports the idea that there is a disconnect between SF and H$_2$ in extreme conditions.

• The morphology of the H$_2$ distribution is also sensitive to the UV field strength and metallicity, with the distribution becoming significantly more compact for lower metallicity and/or higher UV field strength.

• The total gas depletion times for all of our models match those seen in observations of HI-dominated dwarf galaxies, showing that our models compare well with real data.

• The H$_2$ depletion time varies with the metallicity. However, this merely reflects a change in the average H$_2$ mass fraction in the star-forming clouds rather than any substantial change in their ability to form stars. In support of this, we note that the cold gas depletion time shows much less variation than that for the H$_2$.

• A factor of 10 reduction in both metallicity and UV field strength has no significant effect on the star formation rate. However, the star formation rate does differ if these quantities are varied independently of one another.

• The $\Sigma_{SFR}$-$\Sigma_{H_2}$ KS relationship is near linear, but shows a weak dependence on metallicity and UV field strength. $N$ is approximately 1 in the G.10 models, but in the G.01 models we find a steeper slope of $N \sim 1.2$.

### ACKNOWLEDGEMENTS

We thank Paul Clark, Mattis Magg, Ana Duarte Cabral Peretto and the members of the AREPO ISM group for discussions and insightful comments on the coding and science goals in this paper. We also thank the post-graduate groups at Jodrell Bank Centre for Astrophysics for interesting and engaging science discussions and supportive comments, and Volker Springel for access to AREPO. We also thank the referee for the feedback and comments that helped improve the paper. DJW is grateful for support through a STFC Doctoral Training Partnership. RJS gratefully acknowledges an STFC Ernest Rutherford fellowship (grant ST/N00485X/1). SCOG, RT, MCS and RSK acknowledge funding from the European Research Council via the ERC Synergy Grant "ECOGAL – Understanding our Galactic ecosystem: From the disk of the Milky Way to the formation sites of stars and planets" (project ID 855130). They also acknowledge support from the DFG via the Collaborative Research Center (SFB 881, Project-ID 138713538) "The Milky Way System" (sub-projects A1, B1, B2 and B8) and from the Heidelberg cluster of excellence (EXC 2181 - 390900948) "STRUCTURES: A unifying approach to emergent phenomena in the physical world, mathematics, and complex data", funded by the German Excellence Strategy.

This work used the DiRAC COSMA Durham facility managed by the Institute for Computational Cosmology on behalf of the STFC DiRAC HPC Facility (www.dirac.ac.uk). The equipment was funded by BEIS capital funding via STFC capital grants ST/P002293/1, ST/R002371/1 and ST/S002502/1, Durham University and STFC operations grant ST/R000832/1. DiRAC is part of the National e-Infrastructure. The research conducted in this paper used SciPy (Virtanen et al. 2020), NumPy (van der Walt et al. 2011), and matplotlib, a Python library used to create publication quality plots (Hunter 2007).

**DATA AVAILABILITY STATEMENT**

All data is available upon request by the authors of this work.

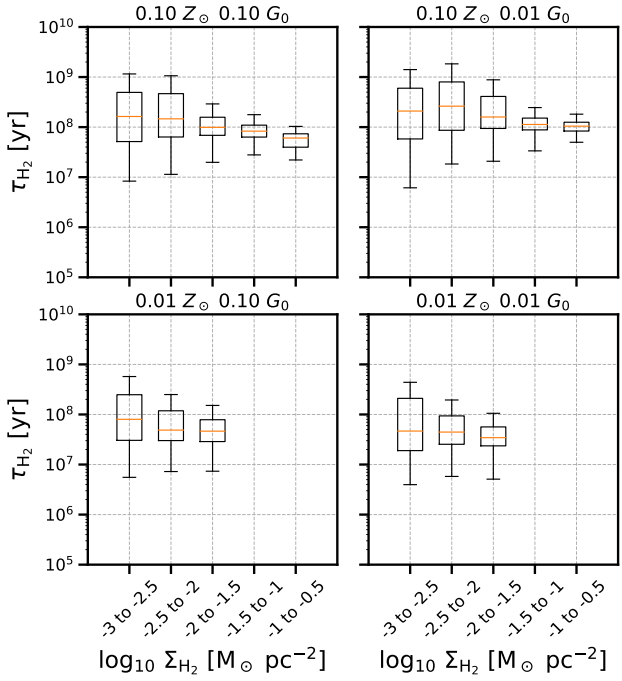

**Figure A1.** Local 1 kpc scale $H_2$ depletion times binned as a function of $H_2$ column density. In all models we see a shorter depletion time in the dense $H_2$ with it rising in the more diffuse gas. We note that there are no $(1\,\text{kpc})^2$ bins with mean $H_2$ surface densities greater than $10^{-2}\,M_\odot\,\text{pc}^{-2}$ in the $0.01 Z_\odot$ models.

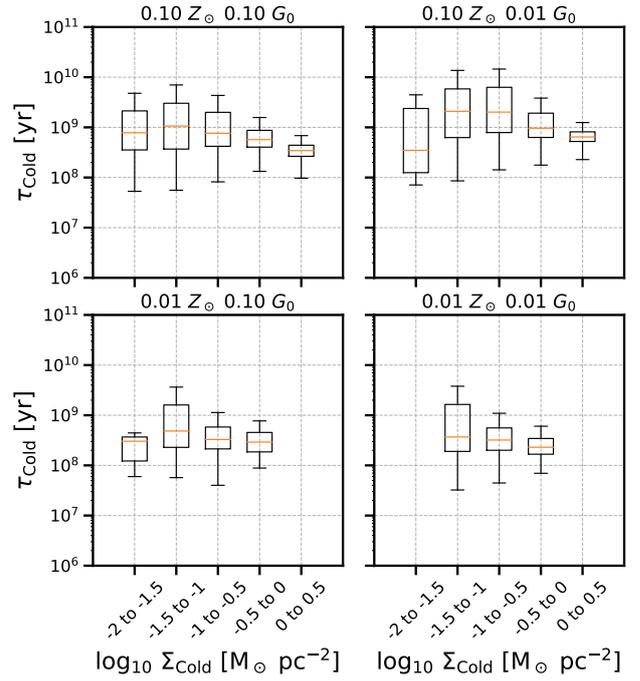

**Figure A2.** Local 1 kpc scale cold gas depletion times binned as a function of cold gas column density. In all models we see a shorter depletion time in the diffuse cold gas with it being fairly consistent across column densities.

|  | $N_{H_2}$ (1 kpc) | $N_{\text{COLD}}$ |
|---|---|---|
| Z.10 G.10 | 1.16 ± 0.014 | 1.26 ± 0.015 |
| Z.10 G.01 | 1.32 ± 0.012 | 1.51 ± 0.015 |
| Z.01 G.10 | 1.45 ± 0.045 | 1.65 ± 0.044 |
| Z.01 G.01 | 1.44 ± 0.046 | 1.59 ± 0.031 |

**Table A1.** Power law slopes (N) for the Kennicutt-Schmidt law for the four models based on $\Sigma_{H_2}$ and $\Sigma_{\text{COLD}}$ at 1kpc resolution with standard errors from linear regression statistics. The KS slope is super-linear for all models with no correlation between models and $H_2$ and cold gas relations.

## APPENDIX A: INFLUENCE OF BIN SIZE ON THE KENNICUTT-SCHMIDT RELATIONSHIP

Here we examine the effects of changing the bin size on the KS relationship that we recover from our simulations. Kruijssen & Longmore (2014) show that a spatial scale of $\sim (500\,\text{pc})^2$ is a good compromise for measuring galactic star formation laws. On smaller scales the relationship breaks down, and moving to larger scales detail is lost by averaging (particularly for a dwarf galaxy). However, different studies do use different apertures. For example, the Bigiel et al. (2008) results are based off a $(750\,\text{pc})^2$ aperture, whilst Pessa et al. (2021) use a range of resolutions from 100 pc to 1 kpc. We run a comparison on the KS relationship by increasing our bin size to $(1\,\text{kpc})^2$ to see what effect this will have on our results.

Our findings are summarised in Table A1. When we increase the bin size, the KS relation steepens, regardless of whether we consider $H_2$ or cold gas, becoming highly super-linear in all models. As before, decreasing the metallicity steepens both slopes, but the impact of varying the UV field strength is more mixed: decreasing it steepens the slope of both relationships in the $0.10 Z_\odot$ simulations, but has no impact on $N_{H_2}$ and actually decreases $N_{\text{COLD}}$ in the $0.01 Z_\odot$ simulations.

If we look at the depletion times for different column densities we see the same trend as for the $(500\,\text{pc})^2$ bins (Figures A1 and A2), with shorter times at higher densities. We also see a downturn in the cold gas depletion time in the lowest cold gas surface density bin, but this appears to be due to the small mass of gas at these densities.

At spatial scales of $1\,\text{kpc}^2$, each bin covers nearly a quarter of the area of the galactic disc, and as we use square bins this includes substantial diffuse gas from outside the disc. As a result of this our KS relationship (for these models at this spatial scale) cannot be considered an accurate representation of the true value. The results we obtain with a $(500\,\text{pc})^2$ bin better follow the gas distribution and therefore are more representative of the true KS relationship within our models.

Finally, to check whether the KS relationship is affected by time evolution, we measured the power-law indices for 3 different time periods in the steady state, taking data once again every Myr but only calculating the indices for periods of 100 Myr, 250 Myr and 500 Myr instead of the entire 700 Myr steady state period. The results can be seen in Table A2. We see small variations when taking

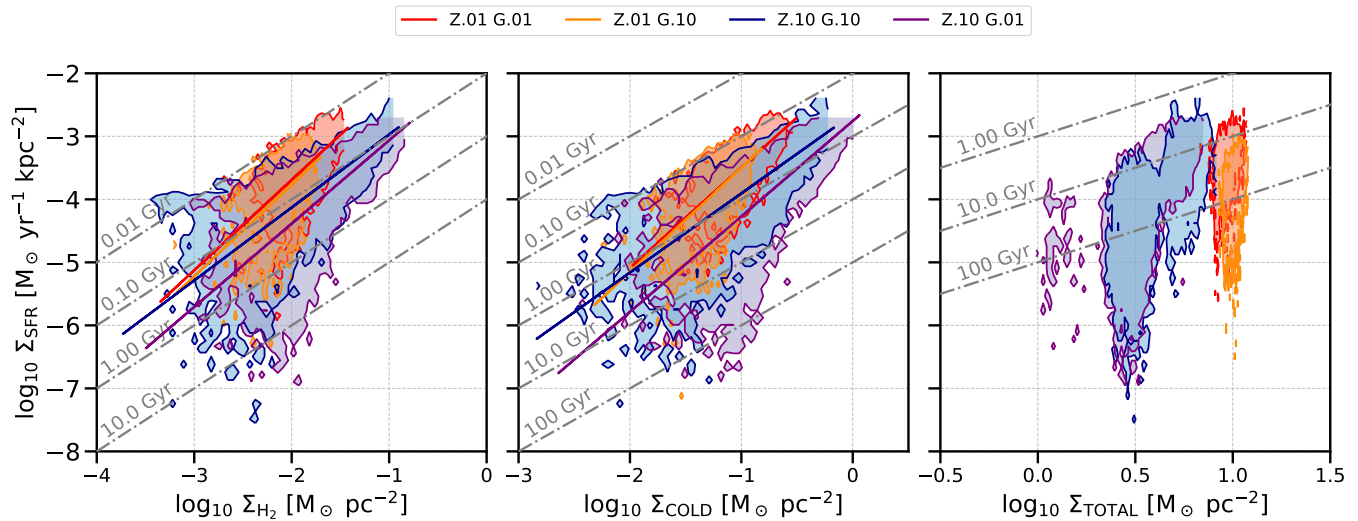

**Figure A3.** Kennicutt-Schmidt relationship for the steady state time period for each model, based on a 1 kpc bin. The legend is the same as the plot for the 500 pc bin. The relationships are steeper than in the 500 pc plot.

data over different time periods but no evidence for large scale time evolution.

# APPENDIX B: ACCOUNTING FOR MOLECULAR GAS IN SINK PARTICLES

In our star formation model, we assume that a fraction $\epsilon_{SF}$ of the gas accreted by sink particles goes on to form stars, with the remaining gas eventually being recycled to the ISM, as explained in Section 2.3. By default, we do not account for this gas when computing the molecular gas or cold gas masses, since we have no definite information regarding its chemical or thermal state. However, it is likely that some significant fraction of this gas is actually cold and molecular. Therefore, in this section we examine how our results would change if we were to assume that all of this gas in sinks younger than 3 Myr[2] was cold and molecular, as in Olsen et al. (2021).

Accounting for this gas results in a slight (∼ 50%) increase in the $H_2$ depletion time, as summarized in Table B1 and Figure B3. However, it still remains substantially smaller than both the cold gas depletion time and the characteristic depletion time of molecular gas in metal-rich spirals. We also find a small increase in the $H_2$ mass fraction, although in each case the gas is still very much HI-dominated.

The molecular KS relationship also flattens slightly when we include the contribution of the gas in the sinks to the molecular gas surface density. This behaviour is expected, since by making this change we are adding additional molecular gas predominantly to regions that have a large number of young sinks, i.e. regions with a high $\Sigma_{SFR}$. However, the change in $N_{H_2}$ resulting from this is relatively small, ranging from approximately zero in run Z.10 G.01 to ∼ 0.1 in run Z.01 G.01. Therefore, none of our results regarding the molecular KS relationship are strongly affected by this change. We also find that the cold gas KS relationship barely changes when we account for the cold gas in sinks, behaviour which is easy to understand if this is only a small fraction of the total amount of cold gas present in our simulated galaxies.

Finally, we show in Figures B2 and B2 how the $H_2$ and cold gas depletion times vary as a function of the corresponding surface density in the case where we account for the gas locked up in sinks. Comparing these with Figures 16 and 17, we see that there is little difference, and hence that our decision in the main text to neglect this gas has little impact on our results.

Given that the inclusion of molecular gas from young sink particles makes the KS relation shallower, rather than steeper, the results in the main portion of the paper can be considered as an upper limit on the steepness of the KS slope with a 500 pc bin. A more detailed sink particle model and analysis would be needed to find the lower limit on the steepness of the KS slope.

This paper has been typeset from a T<sub>E</sub>X/L<sup>A</sup>T<sub>E</sub>X file prepared by the author.

---

[2] We do not consider sinks older than 3 Myr as molecular gas associated with star-forming regions will frequently have been photodissociated or otherwise dispersed by this time; see e.g. Chevance et al. (2020).

|          | $N_{H_2}$ (100 Myr) | $N_{COLD}$ (100 Myr) | $N_{H_2}$ (250 Myr) | $N_{COLD}$ (250 Myr) | $N_{H_2}$ (500 Myr) | $N_{COLD}$ (500 Myr) |
|----------|---------------------|----------------------|---------------------|----------------------|---------------------|----------------------|
| Z.10 G.10 | 1.06 ± 0.037 | 1.19 ± 0.036 | 1.07 ± 0.025 | 1.13 ± 0.024 | 1.05 ± 0.017 | 1.12 ± 0.017 |
| Z.10 G.01 | 1.17 ± 0.034 | 1.19 ± 0.034 | 1.18 ± 0.020 | 1.23 ± 0.021 | 1.17 ± 0.014 | 1.20 ± 0.014 |
| Z.01 G.10 | 1.00 ± 0.066 | 1.25 ± 0.072 | 0.90 ± 0.046 | 1.18 ± 0.049 | 0.96 ± 0.030 | 1.21 ± 0.033 |
| Z.01 G.01 | 1.25 ± 0.069 | 1.42 ± 0.099 | 1.19 ± 0.035 | 1.34 ± 0.034 | 1.25 ± 0.025 | 1.41 ± 0.024 |

**Table A2.** Power law slopes (N) for the Kennicutt-Schmidt law for the four models based on $\Sigma_{H_2}$ and $\Sigma_{COLD}$ at 500 pc resolution with standard errors from linear regression statistics averaged over different time periods. The KS slope is mildly super-linear in almost all models with no correlation between models and $H_2$ and cold gas relations.

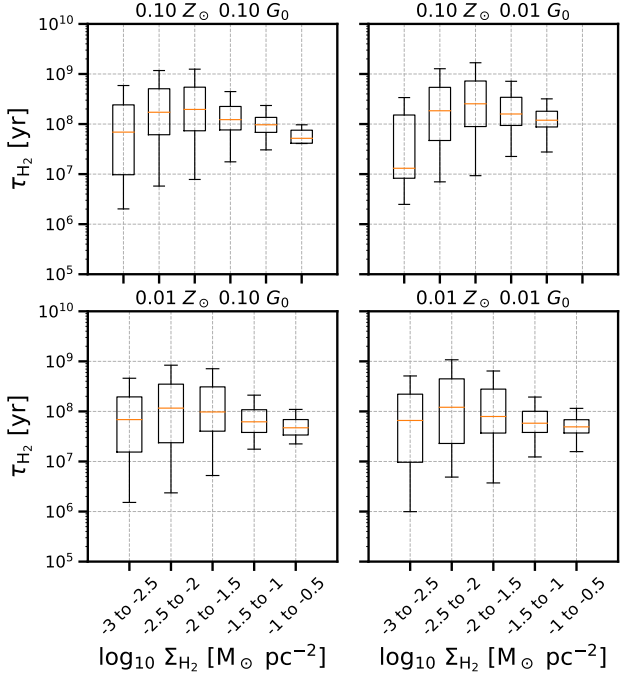

**Figure B1.** Local 500 pc scale $H_2$ depletion times with sinks included binned as a function of $H_2$ column density. In all models we see a shorter depletion time in the dense $H_2$ and an increase in the more diffuse gas.

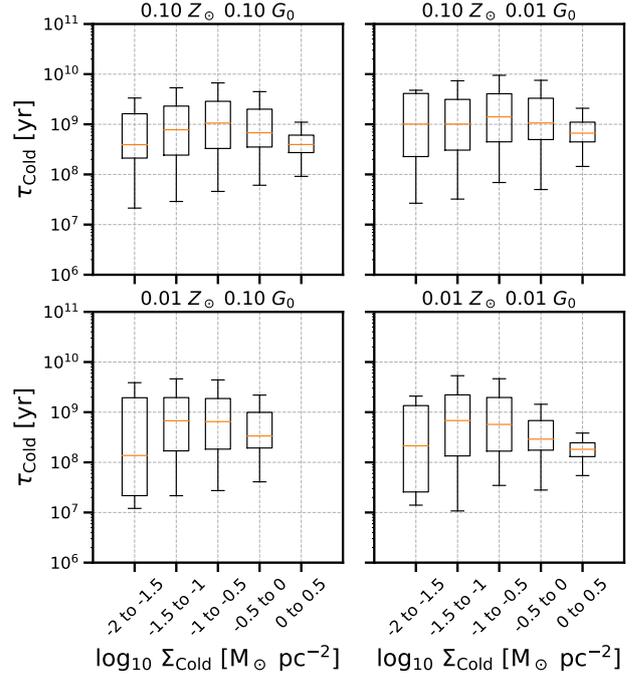

**Figure B2.** Local 500 pc scale cold gas depletion times with sinks included binned as a function of cold gas column density. In all models we see a shorter depletion time in the diffuse cold gas with it being fairly consistent across column densities.

|          | $M_{H_2}/M_{HI+H_2}$ | $\tau_{dep/ss}$ (yr) | $\sigma_{\tau_{dep/ss}}$ |
|----------|----------------------|----------------------|--------------------------|
| Z.10 G.10 | 0.34% | 1.39 ×10$^8$ | 4.53 ×10$^7$ |
| Z.10 G.01 | 0.75% | 1.67 ×10$^8$ | 4.07 ×10$^7$ |
| Z.01 G.10 | 0.06% | 7.88 ×10$^7$ | 2.14 ×10$^8$ |
| Z.01 G.01 | 0.14% | 6.11 ×10$^7$ | 7.77 ×10$^7$ |

**Table B1.** $H_2$ mass fractions and depletion times that we obtain if we assume that all of the non-star-forming gas in sinks 3 Myr old or younger is fully molecular.

|          | $N_{H_2+sinks}$ | $N_{COLD+sinks}$ |
|----------|-----------------|------------------|
| Z.10 G.10 | 1.06 ± 0.014 | 1.15 ± 0.014 |
| Z.10 G.01 | 1.18 ± 0.011 | 1.23 ± 0.012 |
| Z.01 G.10 | 0.88 ± 0.022 | 1.20 ± 0.028 |
| Z.01 G.01 | 1.14 ± 0.018 | 1.41 ± 0.020 |

**Table B2.** Power law slopes (N) for the Kennicutt-Schmidt law for the four models based on $\Sigma_{H_2}$ and $\Sigma_{COLD}$ for a 500 pc bin when we account for the contribution made by molecular gas locked up in young sink particles. We recover slightly smaller values of $N_{H_2}$ compared to the case where we neglect this gas, but $N_{COLD}$ is barely affected.

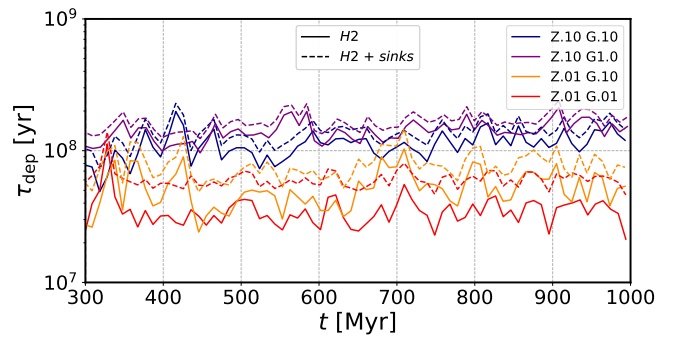

**Figure B3.** Time evolution of the average $H_2$ depletion time when we account for the molecular gas locked up in sinks with ages ≤ 3 Myr (dashed line) and when we do not (solid line). The inclusion of this gas slightly lengths the depletion time in each case.

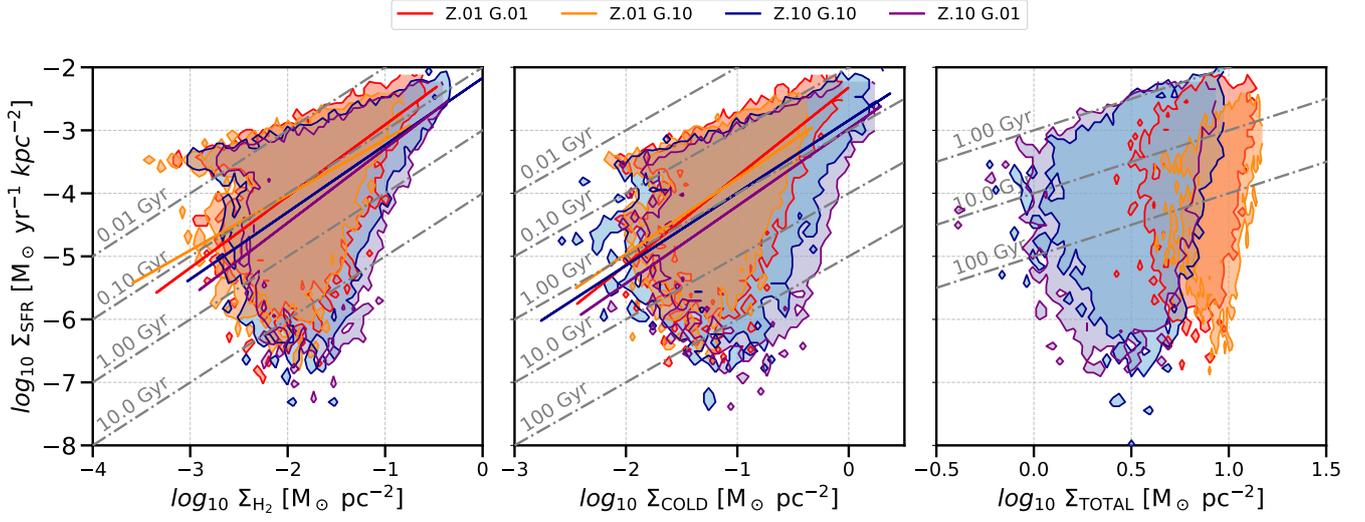

**Figure B4.** Kennicutt-Schmidt relationship for the steady state time period for each model, based on a 500 pc bin and including the contribution to the H$_2$ surface density made by the molecular gas locked up in young (< 3 Myr) sink particles. We see that the inclusion of this gas does not significantly affect the relationship that we recover.

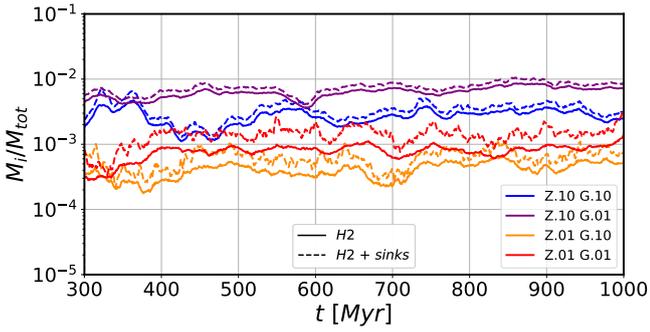

**Figure B5.** As Figure B3, but for the H$_2$ mass fraction. There is a small increase in the mass fraction when we include the contribution from the sinks.